\documentstyle[aps,eqsecnum,epsfig]{revtex}

\newcommand{\ra}{\rightarrow}
\newcommand{\et}{{\em et al.}}

\title{Phase-locking in weakly heterogeneous neuronal networks}
\author{Carson C. Chow} 
\address{Department of Mathematics and Center for BioDynamics, Boston
University, Boston MA 02215}

\begin{document}  
\draft
\date{\today}
\maketitle

\begin{abstract} 

We examine analytically the existence and stability of phase-locked
states in a weakly heterogeneous neuronal network.  We consider a
model of $N$ neurons with all-to-all synaptic coupling where the
heterogeneity is in the firing frequency or intrinsic drive of the
neurons.  We consider both inhibitory and excitatory coupling.  We
derive the conditions under which stable phase-locking is possible.
In homogeneous networks, many different periodic phase-locked states are
possible.  Their stability depends on the dynamics of the neuron and the
coupling.  For weak heterogeneity, the phase-locked states are
perturbed from the homogeneous states and can remain stable if
their homogeneous conterparts are stable. For
enough heterogeneity, phase-locked solutions either lose stability or are
destroyed completely.  We analyze the possible states the network can
take when phase-locking is broken.

\end{abstract}


\section{Introduction}

Phase-locked and synchronized neuronal firing is found throughout the
brain and this coherent activity may play
an important role in behavior and
cognition~\cite{llinas93,gray94,whit95,jeff96,traub96}.  Given their
importance, it 
is of interest to understand the mechanisms responsible for their
generation.  Here, we consider a network of heterogeneous neurons that are
coupled synaptically and determine analytically the conditions under which
their firing patterns will be phase-locked.  Our
results could have implications for synchronous
phenomena in other areas of biology and physics that are based on
pulse-coupled oscillators~\cite{winfree67,strogatz93,herz}.

Much of the previous analytical work on phase-locking and
synchronization in pulse-coupled neuronal oscillators have focused on
homogeneous networks
\cite{mirollo90,vrees94,hansel95,gerstner96,bottani,ernst,vrees96}.
However, any biophysically relevant neuronal network will likely include
heterogeneous effects.  It is known that heterogeneity can greatly
reduce synchrony in networks of phase-coupled
oscillators~\cite{winfree67,kuramoto75,strogatz88}.  Coexistence of
synchronous and asynchronous states have been shown to exist in
large networks of heterogeneous pulse-coupled
oscillators~\cite{golomb93,tsodyks}.  Recently it has 
been shown numerically that heterogeneous assemblies of neurons can
fire coherently. However, the coherent
activity can be greatly reduced even when the heterogeneity
is very small~\cite{wang96,white97}.
Here, we examine analytically the existence and stability of
phase-locked solutions of weakly heterogeneous neuronal networks.  We
consider an all-to-all coupled network of size $N$ where the
heterogeneity is in the intrinsic firing frequency of the component
neurons.  We examine both inhibitory and excitatory connections. 

Phase-locking in homogeneous synaptically
coupled neuronal networks has been studied analytically
with different formalisms~\cite{vrees94,hansel95,gerstner96,terman97}.
Here, we generalize the formalism developed by
Gerstner~\et~\cite{gerstner96} to examine the existence and stability
of phase-locked solutions.  We
find the conditions for stability of homogeneous phase-locked
solutions and examine what happens as we allow the heterogeneity to
increase.  Our formalism can
analyze all periodic
phase-locked states but we mainly focus on the synchronized or
in-phase state, the anti-synchronized state, and the splay-phase or
anti-phase state. 
We find that stable phase-locking depends crucially
on the time course of the synaptic coupling and on the response
properties of the neuron to synaptic inputs.  

Neurons have recently been classified in accordance with their
phase response properties~\cite{hansel95}. Type I neurons  
have the property that their {\em phase response curves} are always
positive, i.e. the next firing time of the neuron is always advanced
in response to a positive (depolarizing) input.  
In contrast, for Type
II neurons the next firing time is either delayed or advanced
depending on when the input is received.  It has also been shown that
Type I neurons admit arbitrarily low frequencies while Type II neurons
do not~\cite{bard96}.
Our results confirm previous observations that for Type I
neurons, excitatory coupling is generally desynchronizing while
inhibitory coupling is
synchronizing~\cite{vrees94,hansel95,gerstner96}.  For Type II
neurons, the opposite can be true although not necessarily.
In previous work, the
requirements on the rise and decay time of the synaptic coupling
for stable phase-locking have been separately 
emphasized~\cite{vrees94,gerstner96,terman97}.  We find that depending
on the nature of the synaptic coupling conditions on both
are required.  A new result for homogeneous neurons is
that the splay-phase or anti-phase state can be stable for 
inhibitory coupling for any finite sized network.  For infinite $N$,
the splay-phase state is known to be unstable~\cite{gerstner95,abbott93}.

With weak heterogeneity, 
phase-locked solutions can exist but with small phase dispersion
around their homogeneous counterparts.  Stability is possible
if the original homogeneous solutions are stable.  The addition of
heterogeneity cannot stabilize an unstable homogeneous phase-locked 
state.   As the heterogeneity is increased, phase-locked
states can either lose
stability or cease to exist.  Heterogeneity can also foster
clustering. 
When stable phase-locking is lost,
the network can enter asynchronous, harmonically locked or suppressed
states. Fig.~\ref{fig:states} shows examples of various states for two
coupled inhibitory neurons obtained from numerical simulations of a
biophysically based neuron model given in Appendix~\ref{app:neu}.

In Sec.~\ref{sec:model} we introduce the model and formalism we use for our
analysis.   In 
Sec.~\ref{sec:lock} we give the conditions which must be satisfied for
a periodic phase-locked solution.  In Sec.~\ref{sec:stab} we give
general conditions for stability of locked solutions. 
In Sec.~\ref{sec:two} we apply our conditions for the existence and
stability of phase-locked states to two coupled
homogeneous and heterogeneous neurons.  We also examine suppression
and harmonic locking.  In Sec.~\ref{sec:N} we examine phase-locking
and other dynamics in a network of N neurons. We discuss our results
and conclude in Sec.~\ref{disc}.

\section{Model}\label{sec:model}

Our network consists of $N$ all-to-all coupled neurons.  Each time a
given neuron fires, a synaptic pulse is transmitted to all the other
neurons.  The effect of the synapse remains over a finite amount of
time.  The
dynamics of neurons are generally described by biophysically based
(Hodgkin-Huxley type) 
equations which are a set of differential
equations for the current balance of the membrane
potential and for the dynamics of the
component ion channels~\cite{rinzel}.  In Appendix~\ref{app:neu}, we
give a set of biophysically based equations~\cite{white97}.  We use
these equations in 
numerical simulations to confirm our analytical results. These neurons
are Type I in contrast to 
classic Hodgkin-Huxley equations which are Type II~\cite{hansel95,bard96}.

\begin{figure}
\centerline{\epsfig{figure=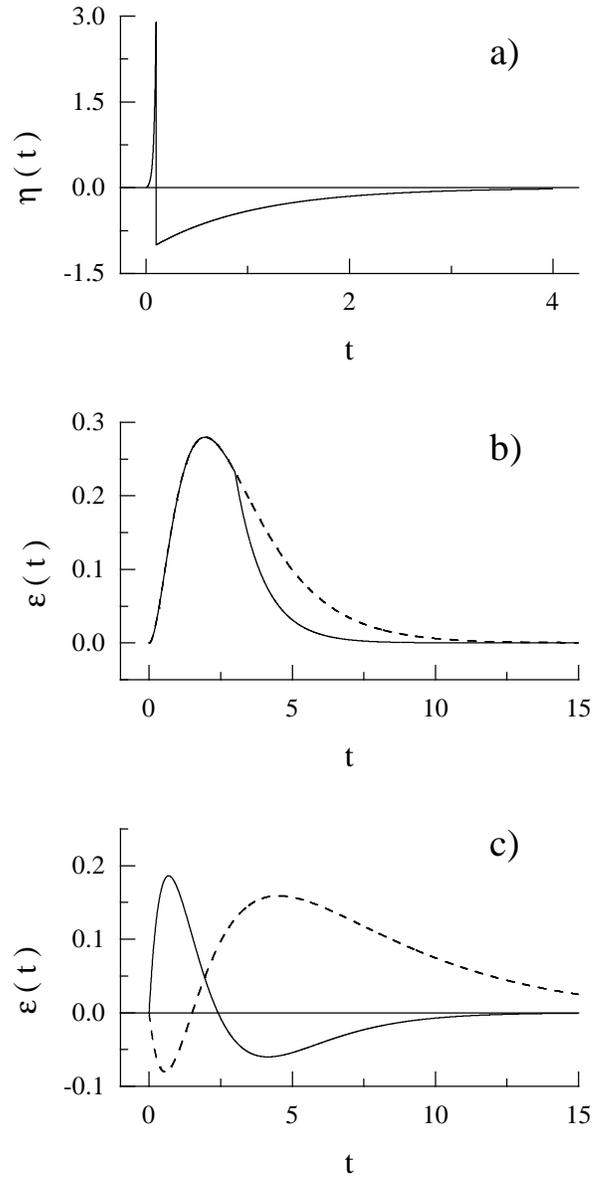,
height=7.in,bbllx=12pt,bblly=12pt,bburx=599pt,bbury=780pt, clip=}} 
\caption{Example kernels. An $\eta$ kernel with a spike is shown in a).
The kernel for the integrate-and-fire model would be 
identical without the spike. In b) we show the $\epsilon$ kernel from
Eq.~(\ref{eps}) 
for the saturating case (solid line) and the nonsaturating case
(dashed line).  Two examples of kernels for Type II neurons are shown
in c).  Note that they have both positive and negative parts.}
\label{fig:kernel}
\end{figure}

Generally, the differential equations for neuron models are
complicated and cumbersome for theoretical analysis.  Instead, we use
the {\em 
spike response method} for our neuronal
dynamics~\cite{gerstner96,gerstner92,gerstner95}.  This formalism is
fairly general and can be applied to biophysical
neuron models.  In 
this method, the 
neuron is considered to be a device which fires when a single
scalar variable (membrane potential) crosses a threshold from below.
The dynamics of the membrane potential consists of a sum of two sets
of kernels akin to an integral expansion:
\begin{equation}
v_i(t) = \sum_l \eta_i(t-t_l^i) +
\sum_{j,l'}\epsilon_{ij}(t-t_{l'}^j),
\label{srm0}
\end{equation}
where $t^i_l$ marks the firing times of neuron $i$ (i.e. the $l$th
time neuron $i$ reaches the threshold $\theta_i$) and
$j$ is summed from 1 to $N$ excluding $i$.  
The kernels are
nonzero only for $t-t_l^k>0$.
The kernel
$\eta_i(t)$ describes the intrinsic dynamics of the neuron after it
has been triggered to fire.  It represents the ensuing spike (action
potential) and recovery behavior.  The kernel $\epsilon_{ij}(t)$
represents the response due to synaptic inputs.  We do not include the
effects of self-coupling in (\ref{srm0}) 
but can do so by modifying $\eta_i(t)$. Figure~\ref{fig:kernel} gives
examples of kernels. 

To linear order we can write $\epsilon_{ij}(t)$ in the form
\begin{equation}
\epsilon_{ij}(t) = \int_0^t G(s) I_{ij}(t-s) ds,
\label{green}
\end{equation}
where $I_{ij}(t)$ is the synaptic input to neuron $i$ from neuron $j$,
and $G(t)$ is the linear response function. It has been shown that the
linear response is often an adequate approximation for the full
nonlinear response~\cite{gerstner96,gerstner95,kistler}.  
In biophysically based neurons, the synaptic input 
current is usually given by $I_{ij}(t)= g_s S_{ij}(t) (v_i-v_s)$ where
$v_s$ is a constant reversal potential, $g_s$ is the maximal synaptic
conductance, and $S_{ij}(t)$ is the synaptic gating variable obtained
from a differential equation (see Eq.~(\ref{synapse}) in Appendix
\ref{app:neu}). Models often use the simplification $I_{ij}(t)\simeq g
S_{ij}(t)$  since $(v_i-v_s)$ is approximately constant away from a 
spike~\cite{hansel95}. The dynamics of the synaptic gating variable
has previously been 
approximated by a recursive scheme where $S_{ij}(t)$ is updated
each time the neuron receives an input
with  $S_{ij}(t) \ra S_{ij}(t) + S_f(t-t^j)$, 
where $t^j$ is the time of the pre-synaptic spike, and $S_f(t)$ is a
stereotypical synaptic response~\cite{vrees94,hansel95,abbott93}.  We
call this type of synaptic model {\em nonsaturating} since $S_{ij}(t)$
increases without bound as the firing rate increases.  On the
other hand the synaptic model given by (\ref{synapse}) is
{\em saturating}. Here $S_{ij}(t)$ saturates to a finite value as the
firing rate of the pre-synaptic 
neuron increases.
In the limit of very fast synaptic rise times we can use 
the approximation $S(t)\ra S_f(t-t^j)$ to model the dynamics of
(\ref{synapse})~\cite{chow97}.  In this case 
the synapse has no memory of previous activity.  

In our network, we allow for heterogeneity in the thresholds of the
neurons which implies heterogeneity in the intrinsic firing
frequencies.  We maintain homogeneity in the time courses of the kernels.
We can make the threshold heterogeneity explicit with the
transformation $v_i\ra v_i-I_i$ and $\theta_i \ra \theta - I_i$, where
$I_i$ can be thought of as a normalized applied driving current, and
$\theta$ is a 
constant threshold.  We make the synaptic strengths of the neurons
explicit by setting $\epsilon_{ij}(t)= J \epsilon(t)$, where $J$
represents the coupling strength.  Heterogeneous coupling could be
accounted for by setting $J=J_{ij}$.  When $J>0$ we call this
excitatory coupling and for $J<0$ we call this inhibitory coupling.
Putting this in Eq.~(\ref{srm0}) yields
\begin{equation}
v_i(t) = I_i +\sum_l \eta(t-t_l^i) + \sum_{j,l'} J
\epsilon(t-t_{l'}^j),
\label{srm}
\end{equation}
In this form, the network heterogeneity is explicitly expressed in the
applied current $I_i$.  We can set the threshold for all the
oscillators to 1 by a simple rescaling without any loss of generality.
In this formulation, $I_{ij}(t)$ in Eq.~(\ref{green}) can be thought of as the
post-synaptic current while
$\epsilon_{ij}(t)$ can be thought of as the post-synaptic potential.
Often we will 
refer to 
the rise time of the synaptic current which will mean the rise time of
$I_{ij}(t)$.  We note that the synaptic kernel $\epsilon_{ij}(t)$ will
itself have a rise time which can depend on both the rise and decay time of
the synaptic current.

\newtheorem{remark}{Remark}
\begin{remark}\label{rem:syn}
{\rm We have assumed that the synaptic kernel does not depend on the
time of arrival or phase of the synaptic input.  For the
integrate-and-fire 
model, as will be shown below, this is true.  However, in general,
there will be a dependence which we can express as $\epsilon_{ij} =
\epsilon(t-t^i_l, t-t^j_m)$. For the ensuing analysis,  we examine
phase-locked solutions where the 
time since the last spike is fixed.  Our assumption is then valid
if $|\partial_u \epsilon(u,v)| << |\partial_v \epsilon(u,v)|$.  For
neurons 
where the response changes drastically with phase, our analysis will
need to be modified.}
\end{remark}

\subsubsection{Integrate-and-fire Model}

To compare to previous analytical results and to provide a simple
concrete example we consider a leaky
integrate-and-fire model~\cite{vrees94,hansel95,gerstner96,tuckwell}
with dynamics given by
\begin{equation}
\frac{dv_i}{dt}=I_i-  v_i -
\sum_l \delta(t-t^i_l) + \sum_{j,l'} J S_f(t-t^j_{l'}),
\label{re}
\end{equation}
where the firing times $t^i_l$ indicate when $v_i$ reaches the
threshold value which we have set arbitrarily to 1.  Each time the
potential reaches threshold a delta function pulse is subtracted to
reset the potential.  As an example for a saturating synaptic current,
consider a double exponential function
\begin{equation}
S_f(t)=\left\{
\begin{array}{ll}\exp(-\beta t)-\exp(-\alpha t),& 0<t<t_{f},\\
                0                              ,&  t_{f}<t,
\end{array}\right.
\end{equation}
where $\alpha>\beta$ and $t_{f}$ is the time elapsed to the next synaptic
input.  For 
nonsaturating synapses, we can set 
$t_{f}=\infty$~\cite{vrees94,hansel95}.

Equation~(\ref{re}) can be integrated
directly~\cite{gerstner96,gerstner95} to obtain the kernels :
\begin{equation}
\eta(t) = -\exp(- t), \quad t>0,
\label{eta}
\end{equation}
\begin{equation}
\epsilon(t)=\left\{
\begin{array}{ll}
\frac{1}{1-\beta}[\exp(-\beta t) - \exp(- t)] -
\frac{1}{1-\alpha}[\exp(-\alpha t) - \exp(-t)],&  0<t<t_f,\\
\left(\frac{1}{1-\beta}\left[\exp(-\beta t_f) - \exp(- t_f)\right] -
\frac{1}{1-\alpha}\left[\exp(-\alpha t_f)  
- \exp(-t_f)\right]\right) \exp(-(t-t_f)),\quad &  t_f<t.
\end{array}
\right.
\label{eps}
\end{equation}
We show examples of the kernels in Fig.~\ref{fig:kernel}.
Note that for $\alpha>\beta$, $\epsilon(t)>0$ so that the time of
the next firing is always advanced in response to a positive
synaptic input and thus the integrate-and-fire model can be 
classified as Type I.

\section{Existence of Phase-locked Solutions}\label{sec:lock}

We first examine the existence of periodic phase-locked solutions.
We derive a 
condition which all periodic phase-locked solutions must satisfy.
Our condition generalizes that of
Gerstner~\et~\cite{gerstner96} to include heterogeneity in the applied
current and locking at arbitrary phases.  Van Vreeswijk~\et~\cite{vrees94}
derived a locking condition for the specific case of two homogeneous
integrate-and-fire neurons locked at arbitrary phases.

Our strategy is to assume
that there is a periodic phase-locked solution and then check for
self-consistency, i.e. the membrane potential must be 1 (the
threshold value) at the time of the next firing.  Consider each neuron
labeled by $i$ to be firing periodically in the past at $t_l^i= (\phi_i-l)T$,
where $l=1, 2,\ldots$, and $0\le\phi_i<1$ is a phase shift.
Inserting these firing times into Eq.~(\ref{srm}) gives
\begin{equation}
v_i(t)=I_i + \sum_{l\ge 1} \eta(t-(\phi_i-l)T) +
\sum_{j\ne i,l'\ge 0} J
\epsilon(t-(\phi_j-l) T), \quad t\le\phi_i T
\label{srm2}
\end{equation}
This assumes
that each neuron will fire once within a period.  We will discuss the
situation where this does not occur later.

Neuron $i$ will fire next at $t=\phi_i T$.  Inserting this into
Eq.~(\ref{srm2}) and imposing self-consistency yields
\begin{equation}
v_i(\phi_iT)=1=I_i + \sum_{l\ge 1} \eta(lT) + \sum_{j\ne i,l'\ge
0} J
\epsilon(l'T+(\phi_i-\phi_j) T)
\label{syncond}
\end{equation}
Note that for $(\phi_i-\phi_j)<0$, the first term of the $\epsilon$
kernel sum will
be equivalently zero by definition of the kernels.  The interpretation is
that within the given period ($l'=0$), neuron $j$ spikes 
after neuron $i$ and the synaptic effect does not appear until the next period,
$l'=1$.
If Eq.~(\ref{syncond}) is satisfied then a phase-locked solution
exists.  The solution will be specified by the set of phases
${\phi_i}$ and the 
period $T$.  We note that the phase locking condition given by
Eq.~(\ref{syncond}) consists of $N$ equations for $N$ phases and the
period.  However, the system is translationally invariant in time so
one of the phases can be arbitrarily fixed.  This then leaves us with
a well defined system of $N$ equations and $N$ unknowns ($N-1$ phases
and the period $T$).

For a homogeneous system, the equations are symmetric with respect to the 
neuron index, implying a large number of possible solutions
for the phases.  The most notable solutions are the in-phase
\cite{winfree67,mirollo90,vrees94,hansel95,gerstner96,kuramoto75} or
synchronized state where all the phases are the same and the
splay-phase state
\cite{aronson,swift,strogatz,schwartz} where the
phases are spaced evenly apart over one period.  There are also
clustered solutions where equal numbers of neurons will fire at evenly
spaced phases.  An example is the anti-synchronous
\cite{vrees94,peskin75,bard84,kuramoto84,abbott90,wang92} 
state where half the oscillators fire a half a period apart.  The
inclusion of heterogeneity breaks the permutation symmetry.  However,
for small enough heterogeneity, near synchronous, anti-synchronous and
splay-phase solutions that are perturbed from their homogeneous
counterparts  can exist. 

\subsection{Phase-locked solutions for two neurons}\label{sec:2lock}
As a specific example, we consider the existence of periodic
phase-locked solutions for the case of two coupled neurons.
We consider neurons with heterogeneous applied current $I_1\ge I_2$ and
homogeneous coupling.
We suppose the neurons have fired at times
$t_l^i= (\phi_i-l) T$, where $l=1,2,3,\ldots$, $i=1,2$.  We choose the
origin of time so that $\phi_2\ge\phi_1$, i.e. neuron 2 always fires
after (or with) neuron 1. 

Applying condition (\ref{syncond}) for
$N=2$ yields
\begin{eqnarray}
v_1(\phi_1 T)&=&1=I_1+\sum_{l\ge 1} [\eta(lT) +J \epsilon(lT-\phi
T)],
\label{vint1}\\ 
v_2(\phi_2 T)&=&1=I_2+\sum_{l\ge 1} \eta(lT) + \sum_{l\ge
0} J\epsilon(lT+\phi T)],
\label{vint2},
\end{eqnarray}
where $\phi=\phi_2-\phi_1\ge 0$ and we have taken into account that
neuron 2 fires after neuron 1.  We must solve Eqs.~(\ref{vint1}) and
(\ref{vint2}) for the phase $\phi$ and period $T$.

A relation for the phase can be derived by subtracting
Eq.~(\ref{vint2}) from (\ref{vint1}) to give
\begin{equation}
0=I_1-I_2 - G(\phi)
\label{Grel}
\end{equation}
where
\begin{equation}
G(\phi) = J \sum_{l\ge 1}[\epsilon(lT-T+\phi T) -
\epsilon(lT-\phi T)].
\label{G}
\end{equation}
For homogeneous neurons, ($I_1=I_2$), condition~(\ref{Grel}) for the
phase is equivalent to that derived by Van Vreeswijk
\et~\cite{vrees94}.  For $\epsilon(0)=0$ (which is generally true), we
find that 
$G(0)=G(0.5)=G(1)=0$.  In addition, $G(\phi)$ is symmetric around $G(0.5)$.
This implies that for homogeneous neurons $\phi=0$ and $\phi=0.5$ are
phase-locked solutions which we refer to as the synchronized and
anti-synchronized solutions respectively.  If additional phase-locked
solutions exist they must always appear symmetrically in pairs on
either side of $\phi=0.5$.  From Eq.~(\ref{Grel}), it is apparent that
neurons with heterogeneous $I_i$ cannot lock in a purely synchronous
or anti-synchronous state. However, if the heterogeneity is not too
large, there can still be phase-locked solutions near to synchrony and
near to anti-synchrony.   For
heterogeneous neurons it is also possible that phase-locked solutions
do not exist at all.  As a specific example, consider the integrate-and-fire 
model. Using 
$\epsilon(t)$ without
saturation from (\ref{eps}) in condition~(\ref{G}), we find
\begin{equation}
G(\phi)= \frac{J}{1-\alpha}\left[
\frac{e^{\alpha\phi T-\alpha T}-e^{-\alpha\phi T}}{1-e^{-\alpha T}}
- \frac{e^{\phi T-T}-e^{-\phi T}}{1-e^{-T}}\right] -  
\frac{J}{1-\beta}\left[
\frac{e^{\beta\phi T-\beta T}-e^{-\beta\phi T}}{1-e^{-\beta T}}
- \frac{e^{\phi T-T}-e^{-\phi T}}{1-e^{-T}}\right] 
\label{Gfunc}
\end{equation}
Figure~\ref{fig:Gphi} shows some examples of $G(\phi)$ for various, $T$,
$\alpha$ and $\beta$.

To specify the period we add Eq.~(\ref{vint2})
to (\ref{vint1}) and divide by 2 to obtain
\begin{equation}
1=\bar{I} + \sum_{l\ge 1} \{\eta(lT) 
+ \frac{J}{2}[\epsilon(lT-\phi T) + \epsilon(lT-T+\phi T)]\},
\label{Trel}
\end{equation}
where $\bar{I}= (I_1+I_2)/2$.
Using Eq.~(\ref{Trel}) and assuming the rise time of the synaptic
current is very short
compared to the decay time ($\alpha>>\beta$), we find that the period
for the 
synchronized state ($\phi=0$) for the integrate-and-fire model with
saturating synapses satisfies~\cite{chow97}
\begin{equation}
1 =\bar{I}(1-e^{- T})+\frac{J (e^{-\beta T} - e^{- T})}{(1
-\beta)}.
\label{trans}
\end{equation}
The period for nonsaturating synapses is given by Eq.~(\ref{trans})
with an extra factor of $(1-e^{-\beta T})$ in the denominator of 
the second term.
The period does not change much for near synchrony where the phase
is near zero.  It has been found that (\ref{trans}) can be used to
estimate the period of a synchronized inhibitory network of
biophysically based Type I neuron models~\cite{chow97}.

\section{Stability of phase-locking}\label{sec:stab}

We examine the local stability of
phase-locked solutions for a network of $N$ all-to-all coupled
neurons.  Our formalism generalizes that of Ref.~\cite{gerstner96} to
include heterogeneity in the applied current $I_i$ and for locking at
arbitrary 
phases $\phi_i$.  In this strategy, perturbations around the firing
times of the phase-locked solutions
are checked for growth or decay. 
Based on the perturbation dynamics we obtain sufficient
conditions for stability.  

For neuron $i$ we consider firing in the past at times
\begin{equation}
t_l^i= (\phi_i-l) T + \delta_i(n-l), \quad l=1,2,3,\ldots,
\label{tlp}
\end{equation}
where $\delta_i(n-l)$ is a small perturbation around each firing time
and $n$ is the index for the current perturbation.  We want to examine
whether $\delta_i(n)$ grows or decays as $n\rightarrow\infty$.  We
will show that $\delta_i(n)$ depends on the perturbations of {\em all}
of the previous firing times of {\em all} of the neurons in the
network.  To derive a mapping for $\delta_i(n)$, we insert the
perturbed firing times (\ref{tlp}) into Eq.~(\ref{srm}) to obtain
\begin{equation}
v_i(t)=I_i + \sum_{l\ge 1}\eta(t-(\phi_i-l)T-\delta_i(n-l)) +
\sum_{j\ne i,l'\ge 0}J
\epsilon(t-(\phi_j-l')T-\delta_j(n-l')). 
\label{vpert}
\end{equation}
The neuron will next fire at $t=\phi_iT+\delta_i(n)$. Hence we can
set $v_i(\phi_iT+\delta_i(n))=1$, which implies after linearizing in
$\delta_i(n)$ that
\begin{equation}
v_i(\phi_iT) \simeq 1 - \dot{v}_i(\phi_iT)\delta_i(n),
\label{vprime}
\end{equation}
where the dot denotes the first derivative with respect to time.  We
now set $t=\phi_i T$ in Eq.~(\ref{vpert}).  Using Eq.~(\ref{vprime})
and linearizing in $\delta_k(l)$ we obtain
\begin{eqnarray}
\lefteqn{1-\dot{v}_i(\phi_iT)\delta_i(n)=I_i+\sum_{l\ge 1}[\eta(lT) -
\dot{\eta}(lT)\delta_i(n-l)]}\nonumber\\ 
&& \mbox{} +\sum_{j\ne i,l'\ge 0}J[\epsilon(l'T + (\phi_i-\phi_j)T)
- \dot{\epsilon}(l'T+(\phi_i-\phi_j)T) \delta_j(n-l')].
\label{pertdyn0}
\end{eqnarray}

For an unperturbed spiking history Eq.~(\ref{syncond}) holds. This
allows us to simplify Eq.~(\ref{pertdyn0}).  Solving for $\delta_i(n)$
yields
\begin{equation}
\delta_i(n)=\frac{1}{\dot{v}_i(\phi_i T)}\left[\sum_{l\ge
1}\dot{\eta}(lT)\delta_i(n-l) +
\sum_{j\ne i,l'\ge
0}J\dot{\epsilon}(l'T+(\phi_i-\phi_j)T)\delta_j(n-l')
\right],
\label{pertdyn}
\end{equation}
where
\begin{equation}
\dot{v}_i(\phi_iT)=\sum_{l\ge 1}\dot{\eta}(lT) +
\sum_{j\ne i,l'\ge 0}J\dot{\epsilon}(l'T+(\phi_i-\phi_j)T),
\label{vp}
\end{equation}
Stability is determined from 
Eqs.~(\ref{pertdyn}) and (\ref{vp}).  It is not
immediately transparent since a given perturbation of a neuron depends
on all previous perturbations including those within the same cycle
(i.e have the same index $n$).  However, by turning the dynamics into
a first return map we can obtain a sufficient condition for stability
which we state in the following theorem.

\newtheorem{theorem}{Theorem} 
\begin{theorem} 
If the series (\ref{pertdyn}) can be truncated then a sufficient 
condition for stability of a phase-locked state is that all of the 
coefficients of the series in (\ref{pertdyn}) are positive.  
\label{theorem1} 
\end{theorem}

\noindent {\em Proof:} We first make $\delta_i(n)$ depend only on
previous firings by substituting for $\delta_j(n)$ recursively on the
right-hand side of the mapping (\ref{pertdyn}) whenever
$\phi_i>\phi_j$.  This substitution can be done systematically if we
index the neurons so that they are ordered by phase with
$\phi_1<\phi_2< \cdots <\phi_N$, i.e. neuron 1 fires first, neuron 2
second, and so forth.  With this phase ordering the mapping for
$\delta_1(n)$ is given by Eq.~(\ref{pertdyn}) with $i=1$ and $l'\ge
1$.  Since neuron 1 fires first it only depends on the previous
firings of the other neurons.  To obtain the mapping for $\delta_2(n)$
we must substitute for $\delta_1(n)$ on the right hand side of
mapping~(\ref{pertdyn}) with $i=2$.  We continue this recursive
substitution scheme for all $N$ neurons where for $i=k$ we must
substitute for $\delta_j(n)$ on the right hand side of
Eq.~(\ref{pertdyn}) for $j<k$.  We write the resulting mapping as
\begin{equation}
\delta_i(n)=\sum_{j=1}^N\sum_{l=1}^\infty M_{ij}(l) \delta_j(n-l),
\label{mapping}
\end{equation}
where $M_{ij}(l)$ is obtained from applying the prescription described
above and the sum over $j$ now includes $i$.  The elements of $M_{ij}$
are composed of products and sums of the coefficients in the series
(\ref{pertdyn}).  Thus, if the coefficients are positive then the
elements of $M_{ij}$ are positive.  Proposition 1 then directly
follows from the following lemma. $\Box$

\newtheorem{lemma}{Lemma}
\begin{lemma} 
If the series (\ref{pertdyn}) can be truncated then a sufficient 
condition for stability is given by $M_{ij}(l)>0$.
\label{lemma1}
\end{lemma}

\noindent
{\em Proof:} By assumption the kernels $\eta(t)$ and $\epsilon(t)$
vanish sufficiently quickly so that we can truncate the series in
Eq.~(\ref{pertdyn}).  This allows us to truncate the inner sum in 
Eq.~(\ref{mapping}) at some $l=l_M$.  We can then express the mapping
(\ref{mapping}) in terms of a first return mapping by embedding in a
higher dimension with
\begin{equation}
\vec{\delta}(n) = {\mathcal M} \cdot \vec{\delta}(n-1)
\label{mateqn}
\end{equation}
where in block form,
\begin{equation}
\vec{\delta}(n) = 
\left(\begin{array}{c} {\mathbf \delta}(n) \\ {\mathbf \delta}(n-1) \\ 
{\mathbf \delta}(n-2) \\ \vdots \\ {\mathbf \delta}(n-l_M+1)
\end{array}
\right),
\qquad
{\mathcal M}=
\left(\begin{array}{ccccc}
{\bf M}(1) & {\bf M}(2) & \cdots & {\bf M}(l_M-1) & {\bf M}(l_M) \\
{\bf I} & 0 & \cdots & 0 & 0 \\ 0 & {\bf I} & \cdots & 0 & 0 \\
\vdots & \vdots & & \vdots & \vdots \\
0 & 0 & \cdots & {\bf I} & 0
\end{array}
\right),
\label{matrix}
\end{equation}
${\bf I}$ denotes the $N\times N$ identity matrix, ${\bf \delta}(k)$
is the column vector of length $N$ with elements $\delta_i(k)$ and
${\bf M}(l)$ is the $N\times N$ matrix with elements $M_{ij}(l)$ from
Eq.~(\ref{mapping}).

The dynamics of the perturbations are determined by repeated matrix
multiplications of ${\mathcal M}$.  The phase-locked solution is
stable if all of the eigenvalues of ${\mathcal M}$ have absolute value
less than unity.  For simple enough systems we can calculate the
eigenvalues explicitly (see Appendix~\ref{app:two}).  This is usually
not the case so we need to determine bounds on the modulus of the
maximum eigenvalue $\rho({\mathcal M})$ (spectral radius).  The
spectral radius is less than or equal to any given matrix
norm~\cite{horn}.  In particular $\rho({\mathcal M})\le
\|{\mathcal M}\|_\infty=\max \sum_{j}|{\mathcal M}_{ij}|$. 
From Eqs.~(\ref{pertdyn}), (\ref{vp}) and (\ref{mapping}) we find that
the rows of $\sum_l {\bf M} (l)$ sum to unity, which implies that all
of the rows ${\mathcal M}$ also sum to unity.  Thus we immediately
find that 1 is an eigenvalue with eigenvector $\vec{\delta}=
(1,1,1,\ldots,1)$.  This corresponds to a global uniform shift in time
and is due to the time translational invariance of the dynamics. It
can be disregarded for stability considerations~\cite{gerstner96}.
If the elements of ${\mathcal M}$ are non-negative then $\rho({\mathcal 
M})=\|{ \mathcal M}\|_\infty=1$. The spectral
radius is the unique maximum modulus eigenvalue if ${\mathcal M}$
is a primitive matrix which is satisfied if ${\mathcal M}^m>0$ for some
$m$~\cite{horn}. 
By inspection of ${\mathcal M}$ from (\ref{matrix}), we see that ${\mathcal
M}^m>0$ if $M_{ij}(l)>0$ proving Lemma~\ref{lemma1}. $\Box$

It is important to note that $M_{ij}(l)>0$ is not a necessary
condition for stability.  We make this explicit in the following
corollary.

\newtheorem{cor}{Corollary}
\begin{cor}
Stability is possible even if $M_{ij}$ has negative elements.
\label{cor1}
\end{cor}

\noindent{\em Proof:} If ${\mathcal M}$ is primitive we know that one
eigenvalue is always 1, it is nondegenerate and all the other
eigenvalues are  within the unit disc.  Stability is lost if an
interior eigenvalue 
crosses the unit boundary, which can occur if we allow some of the
elements of $M_{ij}$ to become negative.  Because the eigenvalues are
continuous functions of the elements, we can make some of the elements
negative and still keep the eigenvalues within the unit disc by the 
intermediate value theorem. $\Box$

\begin{remark}\label{wulf}
{\rm Our perturbation scheme generalizes that of
Gerstner \et~\cite{gerstner96} who arrived at the stability
equation~(\ref{mateqn}) for the purely synchronized solution (all
phases equal).  They considered dynamics where $\dot{\epsilon}(0)=0$ so
only previous perturbations affect the current one.    For a homogeneous
network, 
they proved the necessary and sufficient condition for stability is
$J\sum_l\dot{\epsilon}(lT)>0$. In our case with heterogeneity, pure
synchrony is not possible.  We are thus forced to consider locking at
arbitrary phases and use the recursive
substitution scheme above to calculate the mapping for the
perturbations.}
\end{remark}

\subsection{Stability for two coupled neurons}\label{twostab}
As an example we apply the formalism for $N=2$.  Equation
(\ref{pertdyn}) is
\begin{eqnarray}
\delta_1(n)&=&\frac{1}{\dot{v}_1}\sum_{l\ge 1}
[\dot{\eta}(lT)\delta_1(n-l) + J
\dot{\epsilon}(lT-\phi T)\delta_2(n-l)],\label{pert1}\\
\delta_2(n)&=&\frac{1}{\dot{v}_2}\sum_{l\ge 1}
\dot{\eta}(lT)\delta_2(n-l) + \frac{J}{\dot{v}_2} \sum_{l'\ge0}
\dot{\epsilon}(l'T+\phi T)\delta_1(n-l'),\label{pert2}
\end{eqnarray}
where $\phi=\phi_2-\phi_1$.
Since neuron 2 fires after neuron 1, we substitute
$\delta_1(n)$ from Eq.~(\ref{pert1}) into 
Eq.~(\ref{pert2}) to yield
\begin{eqnarray}
\lefteqn{\delta_2(n)=\sum_{l\ge1}\left\{\left[\frac{\dot{\eta}(lT)}
{\dot{v}_2} + 
\frac{J^2\dot{\epsilon}(\phi T)\dot{\epsilon}(lT-\phi T)}{\dot{v}_1
\dot{v}_2}\right]\right. 
\delta_2(n-l)}\hspace{1cm}\nonumber\\
&&\mbox{} + \left.\left[\frac{J\dot{\epsilon}(lT+\phi T)}{\dot{v}_2} +
\frac{J\dot{\epsilon}(\phi T) \dot{\eta}(lT)}{\dot{v}_1\dot{v}_2} \right]
\delta_1(n-l)\right\}.
\label{pert2b}
\end{eqnarray}
Combining (\ref{pert1}) with (\ref{pert2b}) we obtain
\begin{equation}
{\bf M}(l)=\left(\begin{array}{cc} 
\frac{\dot{\eta}(lT)}{\dot{v}_1} & \frac{J\dot{\epsilon}(lT-\phi T)}{\dot{v}_1} \\
\frac{J\dot{\epsilon}(lT+\phi T)}{\dot{v}_2} 
+ \frac{J\dot{\epsilon}(\phi T) \dot{\eta}(lT)}{\dot{v}_1\dot{v}_2} &
\frac{\dot{\eta}(lT)}{\dot{v}_2} +
\frac{J^2\dot{\epsilon}(\phi T)\dot{\epsilon}(lT-\phi T)}{\dot{v}_1 \dot{v}_2}
\end{array}
\right)
\label{M(l)}
\end{equation}
which is to be used in Eqs.~(\ref{mateqn}) and (\ref{matrix}).  The
sufficient 
condition for stability of a phase-locked solution is given by
Lemma~\ref{lemma1} 
which we denote by ${\bf M}(l)>0$.

For the case of two neurons we can also obtain the
necessary condition for stability by examining whether or not one
neuron can remain locked to the other one independently of whether they are
firing periodically.  This type of stability was considered by van
Vreeswijk \et~\cite{vrees94}.  We derive their condition using the
spike response formalism which we state in the following theorem.

\begin{theorem}\label{nec} 
A necessary condition for stability of a periodic phase-locked
solution for two neurons is
$G'(\phi)> 0$, where $G(\phi)$ is given by Eq.~(\ref{G}) and prime
denotes the first derivative with respect to $\phi$.
\end{theorem} 

\noindent {\em Proof:} Suppose that neuron $1$ fires at times
$t^1_l=t_{1-l}$ and neuron $2$ fires at times
$t^2_l=t_{1-l}+\theta+\delta(n-l)$, where $l\ge 0$.  We consider the
situation where neuron $2$ has fired after neuron $1$.  Neuron $1$
will satisfy the condition $v_1(t_1)=1$ while neuron $2$ satisfies the
condition $v_2(t_1+\theta+\delta(n))=1$, from which we derive the 
linearization $v_2(t_1+\theta) = 1 - \dot{v}_2(t_1+\theta)\delta(n)$.
Inserting this
into Eq.~(\ref{srm}) and linearizing in $\delta(n-l)$ yields
\begin{eqnarray}
1&=&I_1 + \sum_{l\ge 1} [\eta(t_1-t_{1-l}) + J\epsilon(t_1-t_{1-l}-\theta)
- J\dot{\epsilon}(t_1-t_{1-l}-\theta)\delta(n-l)],
\label{pv1}\\ 
1-\dot{v}_2(t_1+\theta)\delta(n) &=& I_2 + \sum_{l\ge 1} [\eta(t_1-t_{1-l})
- \dot{\eta}(t_1-t_{1-l})\delta(n-l) + J\epsilon(t_1-t_{2-l}+\theta)].
\label{pv2}
\end{eqnarray}
Subtracting Eq.~(\ref{pv2}) from Eq.~(\ref{pv1}) and canceling terms,
we obtain the condition for synchrony:
\begin{equation}
I_1-I_2=J\sum_{l\ge
1}[\epsilon(t_1-t_{2-l}+\theta)-\epsilon(t_1-t_{1-l}-\theta)],
\label{Grel2}
\end{equation}
and the perturbation mapping:
\begin{equation}
\delta(n)=\frac{1}{\dot{v}_2(t_1+\theta)}\sum_{l\ge 1}
[\dot{\eta}(t_1-t_{1-l}) - J\dot{\epsilon}(t_1-t_{1-l}-\theta)] \delta(n-l),
\label{sum}
\end{equation}
where
\begin{equation}
\dot{v}_2(t_1+\theta)=\sum_{l\ge 1} [\dot{\eta}(t_1-t_{1-l}) +
J\dot{\epsilon}(t_1-t_{2-l}+\theta) ].
\label{vp2}
\end{equation}
Eq.~(\ref{sum}) can be rewritten as
\begin{equation}
\delta(n)=\sum_{l\ge 1} M(l) \delta(n-l),
\label{map2}
\end{equation}
 where
\begin{equation}
M(l)=\frac{1}{\dot{v}_2}[\dot{\eta}(t_1-t_{1-l}) -
J\dot{\epsilon}(t_1-t_{1-l}-\theta)]. 
\label{newM}
\end{equation}
It will be shown in Lemma~\ref{lemma:mat} that a necessary condition
for stability 
is $\sum M(l)< 1$.

Applying this condition leads to
\begin{equation}
\frac{\sum_{l\ge 1} \dot{\eta}(t_1-t_{1-l})-
J\dot{\epsilon}(t_1-t_{1-l}-\theta)} 
{\sum_{l\ge 1}\dot{\eta}(t_1-t_{1-l})+
J\dot{\epsilon}(t_1-t_{2-l}+\theta)} < 1, 
\label{cond}
\end{equation}
which implies
\begin{equation}
J\sum_{l\ge 1} [\dot{\epsilon}(t_1-t_{1-l}-\theta) +
\dot{\epsilon}(t_1-t_{2-l}+\theta)] > 0,
\label{cond1}
\end{equation}
For periodic firing we can set $t_l= (1-l) T$ and $\theta = \phi T$ in
(\ref{cond1}) which is equivalent to  $G'(\phi)/T>0$ which leads to
$G'(\phi)>0$ as stated in Theorem~\ref{nec}. $\Box$

\begin{lemma}\label{lemma:mat}
For perturbed dynamics of the form
\begin{equation}
\delta(n) = \sum_{l=1}^{l_M} M(l) \delta(n-l),
\label{app1}
\end{equation}
a necessary condition for stability is that $\sum M(l) \le 1$. If
$M(l)>0$, for all $l$, then the condition is sufficient as well.
\end{lemma}

\noindent {\em Proof:} The dynamics can be expressed in matrix
notation as in Eq.~(\ref{mateqn}) where ${\bf M}(l)$ is replaced by
the scalar $M(l)$ and the identity matrices are replaced by 1.  The
corresponding matrix ${\mathcal M}$ is a companion matrix and the
characteristic polynomial is given by~\cite{gerstner96,horn}
\begin{equation}
1=\sum_{l=1}^{l_M} M(l) \lambda^{-l}.
\end{equation}
Let $f(\lambda)=\sum_l M(l) \lambda^{-l}$, so that eigenvalues of ${\mathcal
M}$ satisfy $f(\lambda)=1$.  Suppose $\sum_l M(l) > 1$.  Then $f(\lambda)>1$.
However, $f(\lambda) \ra 0$ as $\lambda \ra
\infty$.  Thus since $f$ is a polynomial and hence continuous, 
$f(\lambda)=1$ at some $\lambda>1$, or in other words one eigenvalue will
be greater than 1.  Thus
$\sum_l M(l) \le 1$ is necessary for all
eigenvalues to be less than 1 and hence for stability.  
If $M(l)>0$, then $\sum_l M(l) < 1$ is necessary and sufficient
for stability since $f(\lambda)<1$, for all $\lambda\ge 1$.  Thus
there are no eigenvalues greater than or equal to 1.~$\Box$

\begin{remark}
{\rm
The condition $G'(\phi)>0$ is not a sufficient condition for periodic
phase-locking.  However, if the dynamics are constrained so that
$M(l)$ as given in Eq.~(\ref{newM}) is positive for all $l$, then
$\sum_{l\ge 1} M(l) <1$ is the necessary and sufficient condition for
arbitrary phase-locking.  However, this constraint does not
ensure periodicity.
For instance, it may be possible that two
neurons could be locked together in a higher order periodic or even
an aperiodic orbit.
}

\end{remark}

\section{Two neuron network}\label{sec:two}

We now apply our formalism to examine the existence and stability of
specific phase-locked solutions for two coupled neurons.  We will revisit
the case of two homogeneous neurons and then consider neurons with
heterogeneous applied current but with homogeneous inhibitory and
excitatory coupling.  In Appendix~\ref{app:two} we consider a simple
example for two neurons where only the kernels from the last firing
are kept.  We compute the eigenvalues for stability explicitly and
compare with our results for the general case.

Applying the stability conditions for phase-locked solutions, we make
the following proposition for the stability of two coupled neurons.
Although we treat the spiking period $T$ as a parameter in the
proposition, it is important to note that it is an emergent
time scale that depends on parameters of the network.

\newtheorem{prop}{Proposition}
\begin{prop}
A periodic phase-locked solution for two coupled neurons is
stable if in (\ref{M(l)}) i) $\dot{\eta}(lT)>0$ for $l\ge 1$,  
ii) $J\dot{\epsilon}(\phi T)\ge 0$ 
and iii) $J\dot{\epsilon}(lT \pm \phi T) > 0$, for all $l\ge 1$.  These are
sufficient but not necessary conditions.
\label{prop:lock}
\end{prop}

Proposition~\ref{prop:lock} is established by
analyzing the matrix elements of ${\bf M}(l)$ in (\ref{M(l)}). By
assumption $\dot{v}_1>0$ and $\dot{v}_2>0$. Conditions i), ii)
and iii) then imply that ${\bf M}(l)>0$ proving the proposition by
Lemma~\ref{lemma1}.  Condition i) is generally true for most neuron
models.  Condition ii) is strongly dependent on the neuron type.
Condition iii) is contingent on the neuron type as well as the neuron
parameters, i.e. strength of the applied current, the
synaptic strength and the synaptic decay time.
Gerstner~\et~\cite{gerstner96} proved Proposition~\ref{prop:lock} for the
homogeneous synchronous case ($\phi=0$) for N
coupled neurons where $N$ is large. We discuss this case in
Sec.~\ref{Nsynch}.  Condition 
iii) corresponds to a special case of their locking theorem. 
This puts a condition on
how slow the onset of the synapse can be.  As will be discussed below,
condition ii) was implicitly assumed in their theorem.  Van
Vreeswijk~\et~\cite{vrees94} emphasized condition ii) in their
criterion for stability.

\subsection{Homogeneous Neurons}

The existence and stability of phase-locking for two homogeneous
neurons has been 
analyzed in detail by van Vreeswijk~\et~\cite{vrees94}  and by
Hansel~\et~\cite{hansel95} for excitatory connections.  These analytical
results used phase
reductions or considered integrate-and-fire dynamics.  Our formalism
is applicable to a more general class of neuron models.  In
Sec.~\ref{het}, we consider heterogeneous neurons.
Gerstner~\et~\cite{gerstner96} considered the stability of the
synchronous solution for a large homogeneous network.  We will
relate the stability conditions of Ref.~\cite{gerstner96} to those
of Ref.~\cite{vrees94}.  Periodic phase-locked solutions are given by
Eqs.~(\ref{Grel}) and (\ref{Trel}) for $I_1=I_2$.  As discussed in
Sec.~\ref{sec:2lock}, these include the
synchronous ($\phi=0$) and anti-synchronous ($\phi=0.5$) solutions as
well as a possible third locked solution.

\subsubsection{Inhibitory Coupling}

Here we relate the results
and formalism of van Vreeswijk~\et~\cite{vrees94} to our
formalism. First consider synchronous locking. 
As mentioned, condition i) in Proposition~\ref{prop:lock} is 
satisfied for most neuron models.   For Type I models, $\epsilon(t)$
is always 
positive so condition ii) can only be satisfied if
$\dot{\epsilon}(0)= 0$. 
This condition is equivalent to a nonzero rise time
of the synaptic current which was emphasized by Van
Vreeswijk~\et~\cite{vrees94} as the condition required for synchrony
with inhibition.  In order for synchrony to occur with inhibition, the
effect of the first neuron on the second one cannot be felt
immediately.  They gave an empirical criterion that the synaptic rise
time be slower than the width of the spike.
For certain Type II models where $\epsilon(t)$
is initially negative and then becomes positive $J\dot{\epsilon}(0)$
could be greater than zero (see Fig.~\ref{fig:kernel}).
Condition iii) is easily satisfied for Type I neurons
(which includes the integrate-and-fire model).
Recall that inhibitory coupling is given by $J<0$. Thus,
if $\epsilon(t)$ is positive with a maximum at $t=t_c$ and is monotone
decreasing for $t> t_c$ (which applies to Type I neurons)
then condition iii) in Proposition~\ref{prop:lock}
is equivalent to $T-\phi T>t_c$.  The decay rate of $\epsilon(t)$ is
controlled to some extent by the decay rate of the synaptic current as
well as the intrinsic membrane decay rate (see Eq.~(\ref{eps})).  The
period on the other 
hand is an emergent property of the network which depends on the
applied current, synaptic strength, and synaptic time
scales~\cite{chow97}.  For 
Type I neurons it can be estimated from Eq.~(\ref{trans}).
For certain Type II neurons it may be
possible that condition iii) can never be satisfied so synchronous
solutions would not be possible with inhibition.

Proposition~\ref{prop:lock} gives sufficient conditions for
stability.  By Corollary~\ref{cor1}, synchrony may be possible even
if Proposition~\ref{prop:lock} is 
violated.  To find when locking is not
possible we examine the necessary condition given by
Theorem~\ref{nec}.  From Eq.~(\ref{G}) we note that the necessary
condition for stability
\begin{equation}
G'(0)\equiv JT\dot{\epsilon}(0) + 2 JT\sum_{l\ge 1} \dot{\epsilon}(lT)>0.
\label{Gprime}
\end{equation}
is satisfied for Proposition \ref{prop:lock} as expected.  If 
$\dot{\epsilon}(0) > - 2\sum_{l\ge
1}\dot{\epsilon}(lT)$ then  synchrony is unstable.  Satisfying this
condition depends on
$\dot{\epsilon}(0)$ which is governed by the rise time of the synaptic
current. In order for stability, the neuron must be able to fire
without being immediately and strongly 
inhibited by the synapse. 
This relates to the van Vreeswijk \et~\cite{vrees94} statement that
the rise of the synapse must be slower than the generation of a spike
for stability to be possible.
For the integrate-and-fire model with nonsaturating synapses, a zero
rise time guarantees 
instability.  This can be seen by examining Eq.~(\ref{Gfunc}) where it
can be shown that $G'(0)<0$ if $\alpha\rightarrow \infty$ and $T>0$.
This verifies previous results of local instability of the synchronous
state for zero rise time with inhibition in the integrate-and-fire
model~\cite{vrees94,gerstner96}.

\begin{figure}
\centerline{\epsfig{figure=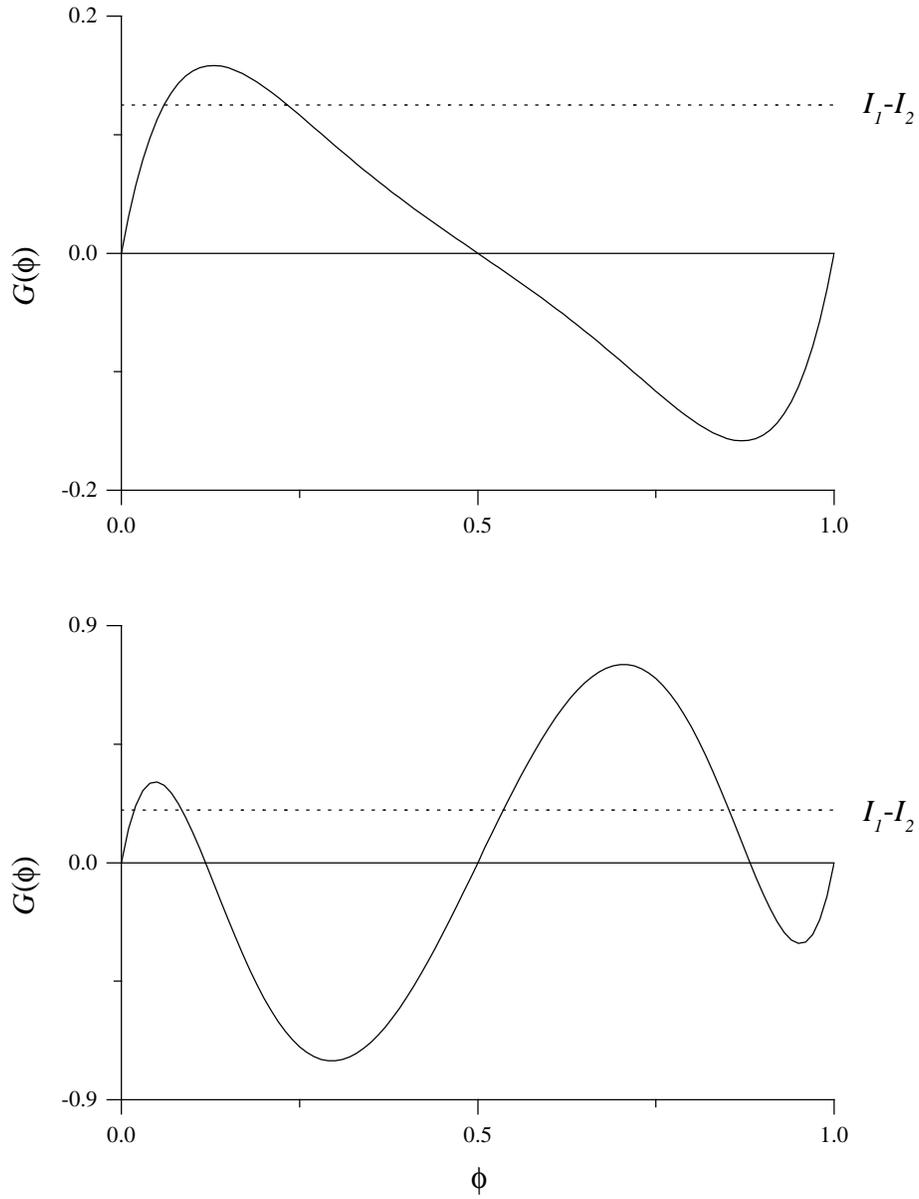, height=7.in,bbllx=12pt,bblly=12pt,bburx=599pt,bbury=780pt, clip=}}
\caption{Two examples of $G(\phi)$ for various values of $\alpha$,
$\beta$, and $T$.  The intersection of $G(\phi)$ and the horizontal
line given by $I_1-I_2$ gives a phase-locked solution.  If $|I_1-I_2|$
is large enough then no phase-locked solution exists.}
\label{fig:Gphi}
\end{figure}

By Proposition~\ref{prop:lock}, the anti-synchronous solution is
stable  as long as the synapse 
is fast enough so that $\dot{\epsilon}(0.5 T + lT)<0$, for $l\ge 0$.
From (\ref{G}) we find the necessary condition for stable anti-synchrony is
given by 
\begin{equation}
G'(0.5)\equiv 2 J T\sum_{l\ge 1} \dot{\epsilon}(lT - 0.5T)>0.
\label{Ganti}
\end{equation}
Condition~(\ref{Ganti}) shows that $G'(0.5 T)$ could be negative if
the onset of inhibition is very slow.  For the integrate-and-fire
model this 
requires that $\alpha$ and $\beta$ be small enough in Eq.~(\ref{eps})
(See Fig.~\ref{fig:Gphi}). This is in accordance with the results of van
Vreeswijk~\et~\cite{vrees94}.

The synchronized and
anti-synchronized solutions can both be stable simultaneously if the
rise time is nonzero and $\epsilon(t)$ is monotone decreasing for 
$t\ge 0.5 T$.  If this is the case then additional
phase-locked solutions must also exist. 
Since $G(\phi)$ is continuous, if $G'(\phi)$ is positive at $0$, $0.5$
and $1$ then it must cross zero at least once between 0 and 0.5 and
between 0.5 and 1.  If it crosses only once then this solution must be
unstable. (See Fig.~\ref{fig:Gphi}).  Since $G(\phi)$ is symmetric
about $\phi=0.5$, we can 
think of these two solutions as one.  This third solution was obtained
numerically in the integrate-and-fire model with nonsaturating
synapses in Refs.~\cite{vrees94,hansel95} where it was found to arise
from a pitchfork bifurcation of the
anti-synchronous solution.  It approaches the synchronous solution as
the synapse becomes faster compared to the period ($\alpha$ and
$\beta$ get larger).  The location of the third solution with respect
to the synchronous and anti-synchronous demarcates the basin of
attractions for the two stable solutions.
All phase-locked solutions can be unstable if
the onset of the synapse 
is so slow that $\epsilon(t)$ is still increasing at $t=T$ because
${\bf M}$ is no longer a positive matrix. 
For slow enough synapses, the necessary conditions for stability given
by (\ref{Gprime}) and (\ref{Ganti}) can both be violated.

\subsubsection{Excitatory Coupling}

The case of slow excitatory coupling for two
neurons has been analyzed previously by van 
Vreeswijk~\et~\cite{vrees94} and Hansel~\et~\cite{hansel95}.  They
both show that for Type I neurons, synchrony is generally unstable and
anti-synchrony is 
stable for slow synapses.  Our analysis agrees with their results.
For two neurons, stability is not ensured since $\dot{\epsilon}(lT)<0$,
violating Proposition \ref{theorem1}.
From Eq.~(\ref{Gprime}) we find that the necessary
condition for synchrony in Theorem~\ref{nec} is violated for
excitatory coupling unless $\dot{\epsilon}(0)$ is positive and large. 
An examination of $G'(0.5)$ for $J>0$ in (\ref{Ganti}) shows that
anti-synchrony is stable if the rise time of the synaptic current
is very slow.

However, stability might be possible for Type II neurons with
excitation as pointed out by Hansel~\et~\cite{hansel95}.  This can
be seen in our formalism because $\dot{\epsilon}(lT)$ could be positive for
$l\ge 1$ if the phase response curve had both positive and negative
parts. We show an example in Fig.~\ref{fig:kernel}.
For the Hodgkin-Huxley equation, Kistler~\et~\cite{kistler}
have shown that the  response of the membrane to a synaptic input can
behave in a way to support synchrony.

For the
integrate-and-fire model with instantaneous (delta function) synaptic
coupling, stable synchrony is possible. To see this consider the
integrate-and-fire 
model (\ref{re}) with a decay constant $\gamma$ for the membrane
potential restored.  This implies that  $\eta(t)=-\exp(-\gamma
t)$ and $\epsilon(t)=\exp(-\gamma t)$.  Thus
$\dot{\epsilon}(t)=-\gamma \exp(-\gamma t)$, which implies
$M_{12}$ in (\ref{M(l)}) is negative.
However, these negative elements can be made arbitrarily small 
if $\gamma$ and $J$ are made small and stability is possible by
Corollary~\ref{cor1}.  For a
combination of small coupling and weak decay, the synchronous state can
be stable.  This is in accordance
with the result of Peskin~\cite{peskin75} who showed that the
product $J\gamma$ must be small for synchrony in the
integrate-and-fire model.  

For the integrate-and-fire model, the third phase-locked solution
which comes from a pitchfork 
bifurcation of the anti-synchronous solution is always
stable~\cite{vrees94,hansel95}.  As the synapse becomes  
faster and more pulse-like the third solution approaches the synchronous
solution.  Thus for very brief synapses, two neurons locked in the
third solution, can be considered
to be in a state of almost synchrony~\cite{pinsky}.

\begin{remark} {\rm It is important to note that
our local stability is different from the global stability
of Mirollo and Strogatz~\cite{mirollo90} and related work
\cite{herz,bottani} for pulse-coupled oscillators.  In these papers, an
absorption condition is assumed for synchrony.
By this it is meant that if two oscillators fire together within a
certain time window, then they are considered synchronized. These
papers show that for pulsatile excitation and almost all initial
conditions, the oscillators 
will eventually get close enough to be absorbed.  What we have
examined here, which is in line with
Refs.~\cite{vrees94,hansel95,gerstner96,peskin75}, is local stability
of coupled oscillators that still influence
each other even if completely synchronized.  After the
oscillators have been absorbed they may not synchronize in the local
sense.  They could be in a state of almost
synchrony, make small oscillations around synchrony, or even be
repelled away~\cite{gerstner96,pinsky}.}
\end{remark}

\subsection{Heterogeneous neurons}\label{het}

We now apply our formalism to two coupled heterogeneous neurons.
In this case,
phase-locked solutions need not exist.  For large enough
heterogeneity or weak enough synaptic strength, condition
(\ref{Grel}) may not be satisfied (See Fig.~\ref{fig:Gphi}). 
When there are no phase-locked solutions, several possibilities can
arise.    If no stable phase-locking exists and both
neurons are able to fire within a period then a possible outcome is a
state of asynchrony, 
where each neuron is effectively decoupled and fires independently of
the other neuron. For strong inhibitory coupling,  we can
get a state of harmonic locking where the neurons are locked at some
ratio of their periods.  The locking period can be very long.
If the inhibition is so strong so that the slower neuron never fires
then we call this state suppression~\cite{white97}.  See 
Fig.\ref{fig:states} 
for numerical examples of these states.  We will discuss harmonic
locking and 
suppression in more detail in the following sections.

Under certain conditions, stable phase-locked solutions can exist. For
weakly heterogeneous
neurons these solutions will be near to the homogeneous locked
solutions.  We concentrate on
the near synchronized and near anti-synchronized solutions.   
From (\ref{Grel}) we see that for a given level of 
heterogeneity,  $|J|$ must
be large enough for a phase-locked solution to exist (See
Fig.~\ref{fig:Gphi}). 
The phase $\phi$ is determined by the
heterogeneity and the coupling strength.  For 
small heterogeneity and small phase, $\phi$ varies linearly with
$I_1-I_2$ and inversely with $-J$.  

For excitatory coupling, the stability results for
homogeneous neurons can be applied to the heterogeneous case.  Near
synchrony is generally
unstable for Type I neurons. For
integrate-and-fire neurons, by analyzing ${\bf M}(l)$ in (\ref{M(l)}),
we see that stability may be 
possible in the case of weak coupling, weak decay of $\eta(t)$, and
extremely brief synapses as in the homogeneous case.  The conditions
for stability of the
near anti-synchronous solution is possible if the
heterogeneity is mild and the synapse is slow.  Near synchrony can
be stable for heterogeneous Type II neurons by
Proposition~\ref{prop:lock}. 

For inhibitory coupling $(J<0)$,
stability of near anti-synchrony can be obtained directly
from Proposition~\ref{prop:lock} but with $\phi$
shifted slightly from 0.5.  As long as conditions ii) and iii) are
still satisfied then 
anti-synchrony is stable.  Stability of 
the near synchronous case where $\phi$ is small but nonzero will
depend on the neuron properties.
If $J\dot{\epsilon}(\phi T) \ge 0$, then stability is ensured by
Proposition \ref{prop:lock}.  For Type I neurons this will not be
generally true but $\dot{\epsilon}(\phi T) = 0$ may hold as a result of
delays in the onset of the synapse or from
refractoriness of the neuron. (We note that $J\dot{\epsilon}(\phi T)<0$
for the integrate-and-fire model~(\ref{re}).) 
For $J\dot{\epsilon}(\phi T) < 0$,
stability is not guaranteed because
elements of the second row of ${\bf M}(l)$ could become 
negative.  However, by Corollary~\ref{cor1}, negative elements do
not immediately imply 
instability because the eigenvalues can remain within the unit circle
even  if ${\bf M}(l)$ takes on negative values.  This leads to the
following proposition on the stability of inhibitory heterogeneous
neurons.

\begin{prop}
Stability of near synchrony for Type I inhibitory coupling is possible
if the heterogeneity and strength of the applied currents are
both weak enough, and the synaptic strength is strong enough.
\label{prop:het}
\end{prop}

We show this by examining the elements of ${\bf M}(l)$ in (\ref{M(l)})
individually and then
determine conditions for them to be positive.  As before, we assume that
$\dot{\eta}(lT)>0$, for all
$l$. We also assume that $\epsilon(t)$ has a single maximum at $t=t_c$
and that $T-\phi T>t_c$ so that
$\dot{\epsilon}(lT\pm\phi T)<0$, for all $l$.  
This implies that 
$M_{11}(l)>0$ and $M_{12}(l)>0$.  
Heterogeneous phase-locking implies
that $\phi> 0$ which affects elements $M_{21}(l)$ and $M_{22}(l)$.

First consider
\begin{equation}
M_{21}(l)=\frac{J\dot{\epsilon}(lT+\phi T)}{\dot{v}_2} 
+\frac{J\dot{\epsilon}(\phi T) \dot{\eta}(lT)}{\dot{v}_1\dot{v}_2}.
\end{equation}
For $J<0$,
this element is positive 
if the magnitude of $\dot{\epsilon}(\phi T)>0$ is small enough and
that of $\dot{\epsilon}(lT+\phi T)<0$ is 
large enough and it
decays away as slow or slower than $\dot{\eta}(lT)$. If we assume that
$\dot{\epsilon}(0)=0$ then $\dot{\epsilon}(\phi T)$ small implies that 
$\phi T$ must be small.  But if $T$ is
too small then $\dot{\epsilon}(lT+\phi T)$ can become too small or
even positive. 
The phase $\phi$ can be made smaller if the level of 
heterogeneity is reduced or if the synaptic strength
$|J|$ is increased. From Eq.~(\ref{trans}) we see that the period $T$
is made longer if the synaptic 
strength is increased or if the applied current is reduced. Thus in
order for $M_{21}(l)$ to be positive we require weak enough
heterogeneity and strength of the applied currents and a strong enough
synapse. 

Next consider
\begin{equation}
M_{22}(l)=\frac{\dot{\eta}(lT)}{\dot{v}_2} +
\frac{J^2\dot{\epsilon}(\phi T)\dot{\epsilon}(lT-\phi T)}{\dot{v}_1
\dot{v}_2}.
\end{equation}
This element is positive
if $\dot{\epsilon}(\phi T)>0$ is small and the decay of
$\dot{\epsilon}(lT-\phi T)<0$ is as fast or faster than 
$\dot{\eta}(lT)$.  Thus requiring $M_{21}(l)>0$  and $M_{22}(l)>0$
implies that 
$\dot{\epsilon}(lT\pm \phi T)$ and $\dot{\eta}(lT)$ 
must have the same decay rates for $l\ge 1$ or there will be negative
elements in the second row of
${\bf M}(l)$ for some $l$. For saturating synapses and fast rise
times, this can be satisfied (see Eq.~(\ref{eps})). The decay rates
will in 
general be different for nonsaturating synapses.  
However, even if this is the case, as long as the decay
rates are fairly close or fairly fast, 
the negative elements can be made arbitrarily small and stability
is still possible by Corollary~\ref{cor1}.
This result indicates that synchronization
should be easier to achieve with saturating synapses as opposed to
non-saturating synapses for heterogeneous
neurons.

\begin{remark}
{\rm 
Proposition~\ref{prop:het} confirms the numerical
observations in Ref.~\cite{white97} where it was found that
synchrony was only observed in what we called the {\em phasic}
regime.  It was shown in Ref.~\cite{chow97} that the 
conditions of Proposition~\ref{prop:het} lead to the phasic regime where
$\beta T \agt 1$. 
We can use this to estimate the boundary between synchrony and
asynchrony in the $J$ - $\beta$ parameter plane.  From our above
analysis and previous
numerical simulations~\cite{white97} we find that $\beta T$ is
approximately constant in the synchronous regime.  At the boundary,
for the level of heterogeneities chosen (5\% in intrinsic frequencies),
we found that $\beta T\sim 1$ defined the boundary.  Recall that the
period of a synchronized 
network with saturating synapses approximately satisfies the
transcendental relation 
(\ref{trans})~\cite{chow97}, which we 
rewrite as
\begin{equation}
J=-\frac{[I(1-e^{- \tau})-1](1-\tau^{-1})}
{(e^{-1}-e^{-\tau})}
\label{gtau}
\end{equation}
where we have substituted $T \sim \tau\equiv \beta^{-1}$.
Eq.~(\ref{gtau}) gives the synchrony-asynchrony boundary for $g$ as a
function of $\beta$.  We can obtain a simpler expression if we let
that $\tau\sim 1+\delta$ and expand for small $\delta$:
\begin{equation}
J\sim -[I(e-1)-e + I\delta.]
\end{equation}
Thus we see that $J$ depends linearly on $\delta\sim\tau-1$.
This was observed
in numerical simulations~\cite{white97}.}
\end{remark}

\subsection{Suppressed States}\label{sec:suppress}
In the previous analysis for phase-locking, it was
assumed that there was a phase-locked
solution with period $T$ and phase $\phi$.  However, for inhibitory coupling
this may not always be true.  
It may be that some neurons never fire at all.
We called such a state suppression in Ref.~\cite{white97}.
We illustrate this for two neurons.  Suppose
$I_1>I_2$ and 
neuron 1 fires before neuron 2.
If neuron 2 never fires (i.e. $\lim_{t\ra\infty} v_2(t)<1$) then we say
neuron 2 is suppressed.  As we will discuss in Sec.~\ref{Nsup},
suppression can also occur in networks of N neurons.

Here we derive the conditions for suppression for two inhibitory
neurons.  We assume that neuron 1
has been firing periodically in the past at $t=(n-l)T$, where $l\ge
1$ and $nT$ denotes the current time.  We also assume that
neuron 2 never fires. We let $J=-g$, where $g>0$.
Using Eqs.~(\ref{srm}) and (\ref{syncond}),
this can be expressed as
\begin{eqnarray}
v_1(nT)&=&1= I_1 + \sum_{l\ge 1}\eta(lT),\label{T1}\\
\lim_{n\ra\infty} v_2(nT) &=& I_2- g \sum_{l=0}^n\epsilon(lT) < 1.  
\label{supcond}
\end{eqnarray}
Equations~(\ref{T1}) and (\ref{supcond}) give the conditions for
suppression. 

Using the kernels~(\ref{eta}) and (\ref{eps}) for integrate-and-fire
neurons with saturating synapses
and using the condition 
$\alpha>>\beta$ we can compute the sum explicitly in
Eq.~(\ref{supcond}) and obtain the condition
\begin{equation}
I_2-\frac{g}{1-\beta}  
\frac{e^{-\beta T}-e^{-T}}{(1-e^{-T})} < 1.
\label{supcond2}
\end{equation}
We see that suppression can be achieved with large enough $g$ or small
enough $T$.  
Without self-inhibition, from Eq.~(\ref{T1}) the period is
given by
\begin{equation}
T=\ln\left(\frac{I_1}{I_1-1}\right)
\label{per}
\end{equation}
If we include the effects of self-inhibition (i.e. the neuron inhibits
itself with strength $g$ when it fires), the period $T$ is given
by Eq.~(\ref{trans}) with 
$\bar{I}=I_1$. 
We can then use Eq.~(\ref{trans}) to simplify Eq.~(\ref{supcond2}) and
obtain 
\begin{equation}
I_2-I_1 + \frac{1}{1-e^{-T}}<1
\label{supcond4}
\end{equation}
From this expression we see that
if the neurons are homogeneous ($I_1=I_2$) then suppression is not
possible.  For heterogeneous cells, if
$I_2<I_1$ then suppression
will definitely occur for large enough $T$. The
longer the period the more likely suppression will occur as opposed to
the case without self-inhibition.  

We can estimate the transition boundary  to suppression in $g-\beta$
space.  Without self-inhibition, the period of neuron 1 is not
affected by the synapse and we can assume that the boundary to
suppression is given by condition (\ref{supcond2}) with an equality
\begin{equation}
I_2-\frac{g}{1-\beta}  
\frac{e^{-\beta T}-e^{-T}}{(1-e^{-T})} \sim 1,
\label{supcond3}
\end{equation}
where the period is given by (\ref{per}).

With self-inhibition, the period now also depends on the synaptic
parameters.  From (\ref{supcond4}) we see that at a critical period
suppression will occur.  The boundary will then be given by the
conditions specifying that period.  Consider 
the case where where $T>>1$ but $\beta T\agt 1$.  This 
corresponds to what we have called the phasic regime~\cite{white97}.  
This allows us to ignore factors of $e^{-T}$ in the period condition
(\ref{trans}).  The 
suppression boundary is then given by 
\begin{equation}
I_1-\frac{g}{1-\beta}  e^{-\beta T}\sim 1,
\label{supcond5}
\end{equation}
for fixed $T$ such that condition (\ref{supcond4}) is minimally
satisfied. 
Solving for $g$ we then obtain
\begin{equation}
g\sim (I_1-1)(1-\beta)e^{\beta T}.
\end{equation}
We see that for $\beta < 1$ and $T$ large enough, $g$ increases as
$\beta$ increases as was observed numerically~\cite{white97}.

\subsection{Harmonic locking}\label{sec:harm}

\begin{figure}
\centerline{\epsfig{figure=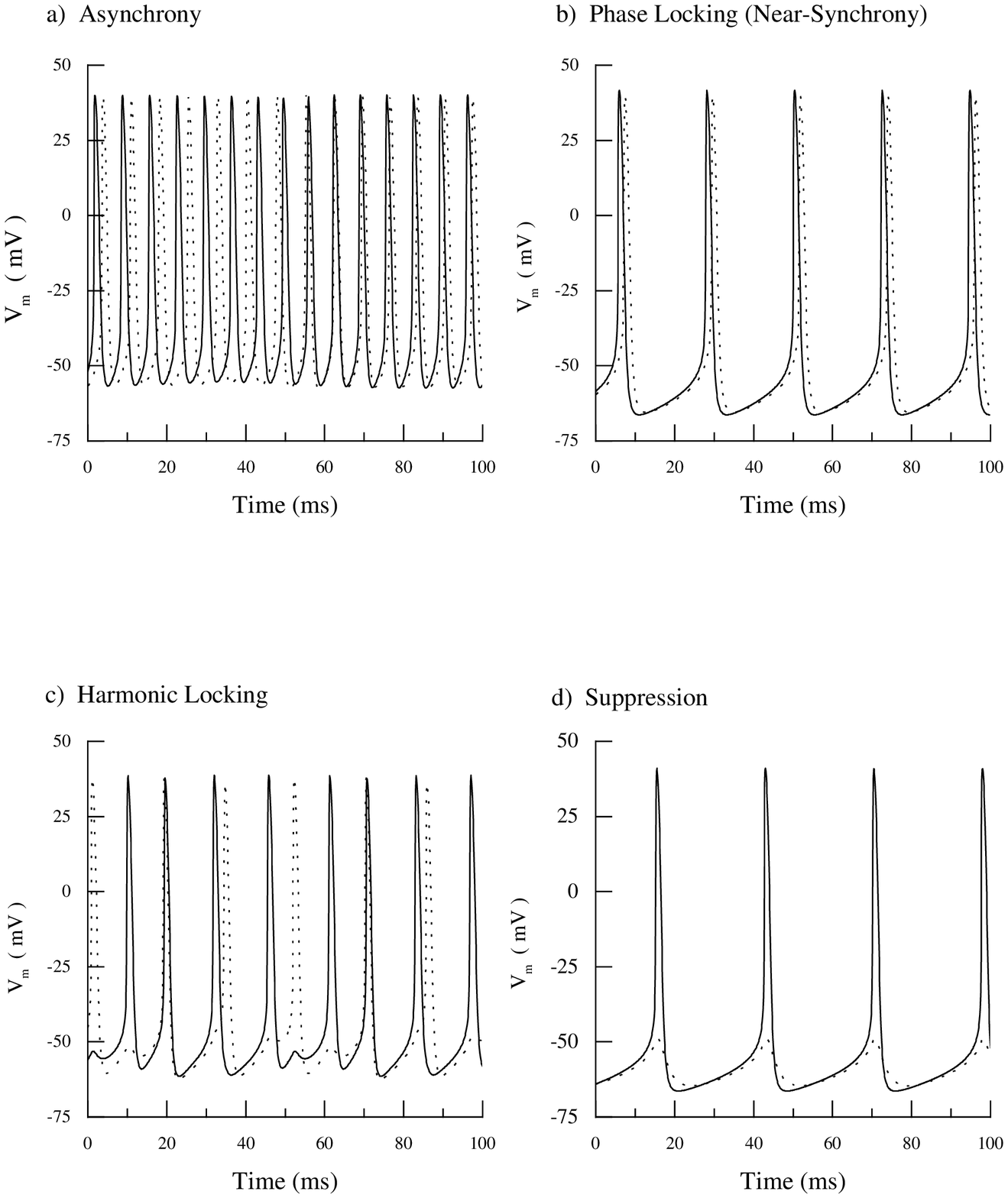, height=7.in,bbllx=12pt,bblly=12pt,bburx=599pt,bbury=780pt, clip=}}
\caption{Examples of possible states for two heterogeneous neurons from
numerical simulations of the model given in Appendix
\ref{app:neu}. 
a)  Asynchrony with $I_1$ = 9.0, $I_2$ = 9.9 $\mu$A/cm$^2$;
$g_s$ = 0.25 mS/cm$^2$; $\tau_s$ = 10 ms.
b)  Near-synchrony with $I_1$ = 1.6, $I_2$ = 1.78 $\mu$A/cm$^2$;
$g_s$ = 0.25 mS/cm$^2$; $\tau_s$ = 10 ms.
c)  Harmonic locking with $I_1$ = 9.0, $I_2$ = 9.9 $\mu$A/cm$^2$;
$g_s$ = 0.5 mS/cm$^2$; $\tau_s$ = 10 ms.
d)  Suppression with $I_1$ = 1.6, $I_2$ = 1.78 $\mu$A/cm$^2$;
$g_s$ = 0.5 mS/cm$^2$; $\tau_s$ = 10 ms.}
\label{fig:states}
\end{figure}

It is also possible that two neurons may be harmonically locked
with a period ratio $p_1:p_2$. The ratio need not be rational.
Suppression can be considered to be harmonic locking at the ratio $1:\infty$.
Consider two neurons locked in the ratio of $p_1:p_2$ so
that neuron $i$ will fire $p_i$ times before repeating its firing
pattern.  Let the
length of this periodic cycle be $T$, which is the 
same for both neurons.
See Fig.~\ref{fig:states} for an example of a $4:3$  harmonically
locked state.  Consider two inhibitory neurons
firing in the past at 
\begin{equation}
t^i_l=  -\left[\frac{l}{p_i}\right]T -\phi^i_{l \bmod p_i}T,
\end{equation}
where $[\cdot]$ indicates least integer, $l=0,1,2,\dots$, 
and $\phi^i_{l \bmod p_i}$ is the phase advance per firing which
repeats after $p_i$ firings. 
In the next cycle, neuron $i$ will fire at
$t=T-\phi_{m_i}^i$, where $m_i$ takes the values $0, 1,\dots,p_i-1$. 
If we substitute this into Eq.~(\ref{srm}) we obtain
\begin{equation}
v_i(T-\phi^i_{m_i})=1=I_i+\sum_{l\ge m_i}
\eta\left(T+\left[\frac{l}{p_i}\right]T + 
(\phi^i_{l \bmod p_i} -\phi^i_{m_i})T\right) - 
\sum_{l'\ge n}g\epsilon\left(T+\left[\frac{l'}{p_j}\right]T + 
(\phi^j_{l'\bmod p_j}  -\phi^i_{m_i})T\right), 
\label{harmlock}
\end{equation}
where $n$
is the index for which $\phi^j_{n \bmod p_j}-\phi^i_{m_i}>0$ is
minimal (i.e. $n$ 
gives the index of the nearest firing of the other neuron).
Equation~(\ref{harmlock}) is a system of $M\equiv p_1+p_2$ equations
that must be 
solved simultaneously for the period $T$ and the phases $\phi^i_{m_i}$.
Because of the freedom to shift globally in time we can fix one of the
phases (e.g. set $\phi^1_0=0$), which then leaves $M$ equations for the
remaining $M-1$ phases and the period $T$.  For a fixed set of
parameters, there is an infinite set of possible harmonic states with
different period ratios.

Stability of the harmonic state can be established by applying results of
the previous section.  A stability analysis can be performed as before
by perturbing the time of each firing and constructing a mapping for the
perturbations.  The mapping will map all the perturbations from one
cycle at $t= k T$ to the next cycle at $t=(k+1) T$, where $k$ is the
least integer of $l/p_i$.  The sufficient
condition for stability will be given by Theorem~\ref{theorem1}.
We can examine the local stability of the relative firing times of the
two neurons.  As shown in Sec.~\ref{twostab}, for inhibitory coupling, a
sufficient condition for 
stable phase-locking is that the phases between two consecutive spikes
be either very small or far apart.  More specifically terms such as
$g\dot{\epsilon}(kT+(\phi^j_{l \bmod p_j}-\phi^i_{m_i}) T)$ will arise in
the perturbation series.  The terms with $k=0$
correspond to the term $g\dot{\epsilon}(\phi T)$ in Sec.~\ref{het}.
Recall that stability required $g\dot{\epsilon}(\phi T)$ to be small.
Terms for $k>0$ will be positive if the synapse decays fast enough.
This then implies that stable harmonic locking will involve spiking
patterns where the spikes from the two neurons are generally either
coincident, close together or fairly far apart (See
Fig.~\ref{fig:states}.  This was observed 
numerically~\cite{white97}.

Numerically, harmonic locking was more likely to be observed if
self-inhibition was included~\cite{white97}.  One reason may be that
the locking conditions (\ref{harmlock}) are easier to
satisfy with self-inhibition because the equations become more
symmetric.  Another reason may be that the transition from synchrony
to suppression is slower as the parameters are changed for
neuronal networks with
self-inhibition compared to those without it, leading to a larger
region in parameter space where harmonic locking is observed. 

\section{$N$ Coupled Neurons}\label{sec:N}

We now examine the dynamics of a network of $N$ neurons.  Many of the
results from two coupled neurons can be carried over to the $N$-neuron
network.  For $N$ homogeneous neurons there are many possible
phase-locked solutions.  As discussed in Sec.~\ref{sec:lock}, all
symmetric combinations of the phases are solutions.
Theorem~\ref{theorem1} gives the condition for stability.
We examine the existence and stability of
synchronous, splay-phase,  clustered, and unlocked states.  Large
homogeneous networks have been examined
previously using different formalisms~\cite{hansel95,gerstner96,vrees96}.  
We consider the effects of weak heterogeneity.
It is known that strong heterogeneity can induce
various complicated states~\cite{golomb93,tsodyks}.

\subsection{Synchrony}\label{Nsynch}

For a homogeneous network, the synchronous state is a phase-locked
solution. Stability has been examined in the limit of large $N$ by
Gerstner \et~\cite{gerstner96} (See Remark \ref{wulf}).
They proved that as long as the influence of the synaptic kernel is
rising at the time of the next spike then synchrony is stable, i.e.
$\sum_{l\ge 1}J\dot{\epsilon}(lT) >0$, where $J<0$ ($J>0$) corresponds
to inhibitory (excitatory) coupling.  Implicit in their proof is that
the synaptic kernel does not act immediately
(i.e. $\dot{\epsilon}(0)=0$).

With the inclusion of heterogeneity, the phases are shifted apart for a
locked solution.
However, for weak heterogeneity, a stable near synchronous solution 
is still possible.  This was observed in numerical
simulations~\cite{white97}.  Here we illustrate that for inhibitory
coupling, if the neurons are indexed so that $I_1>I_2>\cdots> I_N$ then
there could be a stable phase-locked solution where the phases are
arranged as $\phi_1<\phi_2<\cdots<\phi_N$ provided the 
heterogeneity is small and close to uniformly distributed.
We  translate time so that the phases
are all positive and small.

First consider a homogeneous system with a period satisfying
the condition
\begin{equation}
1=\bar{I} + \sum_{l\ge 1} [\eta(lT) -g(N-1) \epsilon(lT)].
\label{Ibar}
\end{equation}
where $J=-g$, $g>0$, and $\bar{I}\equiv (1/N)\sum_j I_j$.  We now let
$I_i=\bar{I}+\Delta_i$.  Since the applied currents are all positive
then $\Delta_1>0$ and $\Delta_N<0$.  If a 
phase-locked solution exists in the heterogeneous system it must
satisfy the condition 
\begin{equation}
1=\bar{I}+\Delta_i + \sum_{l\ge 1}\eta(lT) - \sum_{j\ne i, l\ge 1}
g\epsilon(lT + (\phi_i-\phi_j) T) - \sum_{j<i} g\epsilon((\phi_i-\phi_j)T),
\label{Ii}
\end{equation}
Subtracting condition (\ref{Ibar}) from  (\ref{Ii}) gives
\begin{equation}
\Delta_i = \sum_{j\ne i, l\ge 1} g[\epsilon(lT+(\phi_i-\phi_j) T) -
\epsilon(lT)] +\sum_{j<i}g\epsilon((\phi_i-\phi_j)T)
\label{Di}
\end{equation}
Consider condition (\ref{Di}) for neuron 1 ($i=1$):
\begin{equation}
\Delta_1 = \sum_{j\ne 1, l\ge 1} g[\epsilon(lT+(\phi_1-\phi_j) T) -
\epsilon(lT)].
\label{D1}
\end{equation}
Since we have assumed that $\phi_1<\phi_j$, then
condition (\ref{D1}) can be satisfied if
$\epsilon(t)$ is monotone decreasing for $t>T+(\phi_1-\phi_N)T$.
Now consider condition (\ref{Di}) for neuron $N$:
\begin{equation}
\Delta_N = \sum_{j< N, l\ge 1} g[\epsilon(lT+(\phi_N-\phi_j) T) -
\epsilon(lT)] + \sum_{j<N}g\epsilon((\phi_N-\phi_j)T)
\label{DN}
\end{equation}
In this case $\Delta_N<0$ and $\phi_N>\phi_j$.  Condition
(\ref{DN}) can be satisfied if $\epsilon(t)$ is monotone
decreasing for $t>T+(\phi_N-\phi_{N-1}) T$ and
$\epsilon((\phi_N-\phi_j)T)$ is not too 
large. We can perform a similar analysis for all the other neurons.
The result is that
as long as $\epsilon(t)$ is monotone decreasing for
$t>T+(\phi_1-\phi_N)T$ and it does not rise up too
quickly for small $t$ then near synchrony is possible with the phases
arranged as $\phi_1<\phi_2<\cdots<\phi_N$, i.e. they 
will all fire in rapid succession.  We should note that the period is
controlled by the applied currents, the synaptic strengths, and the
synaptic time course. 

Stability is determined from the perturbation
series for stability (\ref{pertdyn}) which will be composed of terms
with coefficients of 
the form $\dot{\eta}(lT)$, $-g\dot{\epsilon}(lT +(\phi_i-\phi_j)T)$,
and $-g\dot{\epsilon}((\phi_i-\phi_j)T)$, for all $i$, $j$ and $l$.
By Theorem~\ref{theorem1}, the phase-locked solution is stable if all
of the coefficients are positive.  The monotone decreasing condition
guarantees that the first two sets of coefficients are positive.  For
a homogeneous system, the third set is zero.  For a heterogeneous
system, the third set can be negative for Type I neurons.  However,
by constructing the 
return mapping (\ref{mateqn}) it can be shown that stability is
possible as long as they are small by using Corollary~\ref{cor1} and
generalization Proposition~\ref{prop:het}
to the  heterogeneous N neuron network.

In the case of Type I excitatory coupling and zero rise time of the
synaptic current, synchrony can be globally stable but it can be locally
unstable~\cite{gerstner96}. If a neuron fires
late with respect to the 
others then it will be pushed forward and eventually catch up to the
synchronous 
population, but if it fires 
early it will be pushed away until it catches up to the next cycle.
Mirollo and Strogatz~\cite{mirollo90} 
considered an absorption condition so that once two neurons fire
within a given time window they are considered
synchronized and remain synchronized.  This absorption condition
ensures global stability.   

\subsection{Splay-phase and Clustered States}

For a homogeneous network, the splay-phase state  also known as
anti-phase, `rotating wave' or the more colorful `ponies on a
merry-go-round'
\cite{hansel95,vrees96,aronson,swift,strogatz,schwartz}
is a possible phase-locked solution.  In this state, all $N$ neurons
in the network fire with a  phase separation of $1/N$. This solution can be
verified by 
examining condition (\ref{syncond}).  With weak heterogeneity, a near
splay-phase state should exist by a similar argument to the perturbed
synchronous case.

We can show stability of the splay-phase state with an
$N$ neuron version of Proposition~\ref{prop:lock}. We note that the
coefficients of (\ref{pertdyn}) are composed of factors of the form
$\dot{\eta}(lT)$ and $J\dot{\epsilon}(lT\pm j\phi T)$ where $\phi= 1/N$,
and $j=1,\dots, N$ is the neuron index.
If these factors are positive then there is stability by Theorem
\ref{theorem1}.  
In the excitatory case, the
splay-phase solution can be stable if the rise time of the synapse is
very slow so that $\dot{\epsilon}(j\phi T)>0$.  In this case the
splay-phase is stable for similar reasons as to  why the
anti-synchronous solution is stable for two neurons.  This has been
noted previously~\cite{hansel95,vrees96}.  With weak heterogeneity,
stability of the near splay-phase state can still be stable provided
the synapse is slow enough.

The splay-phase state can also be stable for Type I inhibitory
connections if  
the synapse has a fast enough rise time so that $\dot{\epsilon}(T/N)<0$
and $\epsilon(t)$ is monotone decreasing for 
$t> T/N$.  Stability may be affected if the action potentials (spikes)
of the individual neurons overlap in the splay-phase state.  This can
occur if the 
network is so large that $T/N$ is on the
order of the spike width.  
We have verified the existence of the splay-phase state
numerically in our biophysical model 
for $N=4$.  The stability of the splay-phase is interesting because
in the mean field limit ($N\ra\infty$) it has been
shown to be unstable for inhibitory coupling
\cite{abbott93,gerstner95,treves}.  In the mean field limit for 
homogeneous neurons, the splay-phase state corresponds to an
`asynchronous' state where the density of the firing times of the
neurons is constant.  We can see how this instability arises in our
formalism. As more neurons are added, the synapse must rise 
faster  for stability.  In the limit of 
infinite numbers of neurons, stability is not possible no matter how fast
the synapse since $(T/N)\rightarrow 0$.
With weak heterogeneity, a stable perturbed splay-phase state is
possible as long as the rise time of the synapse is shorter than the
smallest phase 
difference.  Stability should be more
affected by heterogeneity as the network size increases since
the phase differences become smaller.

There is also the possibility of phase-locked
clustered states.
Clustering in neuronal networks has been investigated 
before~\cite{hansel95,vrees96,golomb92,golomb94,kopell94}.  For
homogeneous 
networks, groups of neurons can fire
together in unison but at different phases with other groups.
Clustered states can be shown to be stable if $\dot{\epsilon}(0)=0$ and
$J\dot{\epsilon}(l T+(\phi_a-\phi_b)T)>0$, where $a$ and $b$ are the
indices 
for different groups.  There are many possible coexistent
clustered states that depend on the initial conditions.  With
heterogeneity, clustered solutions of near synchronous neurons are
possible as in the near synchronous solution.  If the
heterogeneity is nonuniform and clustered then groups may form more
easily among neurons with similar intrinsic frequencies.

\subsection{Asynchrony, Harmonic Locking and Suppression}\label{Nsup}

Heterogeneity breaks the symmetry of the
network.  As we have seen above, if the heterogeneity is mild then
phase-locked solutions can exist.  For large enough heterogeneity
there are no phase-locked solutions.
When phase-locking does not exist several possibilities can arise.  We
have discussed asynchrony, harmonic locking and suppression for two
coupled neurons.  These states have their counterparts for a network
of N neurons as well. 

If the heterogeneity is large then our analysis in Sec.~\ref{Nsynch}
shows that stable synchrony can be broken.  If the applied currents
are very strong or the synapses are very weak then the period can become
too short to support stable synchrony.  In this case, the neurons will
become effectively decoupled and fire asynchronously as in the two
neuron case. This was observed numerically~\cite{white97}. For
heterogeneous neurons, the average density of neurons at a given phase
need not be constant as in the homogeneous case.
If the heterogeneity is very broad then it is possible that asynchrony
and synchrony can coexist~\cite{golomb93,tsodyks}. 

If the applied currents are weak and the synapses are strong then
stable synchrony can again be broken.  For inhibition, the possible
outcomes are harmonic locking or suppression.  Here the slower neurons
cannot keep up with the faster ones.  The dynamics will be similar to
two self-inhibited neurons.  The influence of
the other neurons in the network with weakly heterogeneous applied
currents acts like self-inhibition.  If a 
neuron is too slow compared to a group of other neurons, then it can
drop out of the rhythm.  As the synaptic strength is increased, more
and more neurons will drop out until only one remains.
The condition for suppression can be
estimated from the combined 
synaptic input the neuron receives over a period.  Similarly, complicated
harmonic patterns can form where the slower neurons lock to sub
harmonics of faster neurons.  In this case a large set of coexistent
patterns exist~\cite{white97}.

\section{Discussion and Conclusions}\label{disc}

We have examined the existence and stability of phased-lock solutions
in a network of neurons with heterogeneous intrinsic frequency.  With mild
heterogeneity, we have shown that stable phase-locking is possible but
it is fairly fragile.  When periodic phase-locking breaks, states of
asynchrony, harmonic 
locking or suppression may arise.  In accordance with previous
work~\cite{vrees94,hansel95,gerstner96},
we have found that phase-locking tendencies are strongly determined by
the type (I or II) of the neuron
model.  We also distinguish between
 saturating and nonsaturating
synaptic types and find 
that there can be differences in phase-locking
properties.  Neurons
with saturating synapses tolerate the effects of heterogeneity
better. 

Our analysis may clarify some of the apparent differences in previously
derived conditions for synchronization with inhibition.  We find that stable
synchrony requires a nonzero rise time of the synaptic current as
emphasized in Refs.~\cite{vrees94,terman97}.  This was implicitly
assumed in Ref.~\cite{gerstner96}.  We also find that it requires
that the rise time not be too slow which was emphasized
in Ref.~\cite{gerstner96} and assumed in Ref.~\cite{terman97}.  
However,
Ref.~\cite{terman97} also requires that the synaptic decay time
not be too fast which was also observed in numerical
simulations in Ref.~\cite{wang92}.  This condition may be implicit in
the behavior of the kernels.  For some neuron models, the
$\epsilon$ kernel may become negative if the synaptic decay time
becomes too fast.  For example, if the threshold of the
integrate-and-fire model changes with the synapse then this condition
may be present~\cite{terman97}.

Our results rely on the spike response method to characterize the
dynamics of a threshold crossing variable in terms of a set of
response kernels.   
The existence and stability of phase-locked solutions are 
then established by the time courses of these kernels.
Critical to
our analysis is that the kernels exist and can be computed. 
For the integrate-and-fire model the kernels can be obtained
explicitly.  For more complicated models, they most likely must be
obtained numerically~\cite{kistler}.  However, we can still infer
general properties on how they should behave.  

The neuron type (I or II) governs the general
characteristics of the 
synaptic kernels.  For example, positive inputs always advance the firing
of Type I neurons.  This implies that the synaptic kernel
always remains positive.  
As noted previously~\cite{hansel95,bard96}, we find that
Type I and Type II synapses will greatly affect the phase-locking
tendencies of the neurons.  Our formalism shows why this is
the case from the shapes of the synaptic kernels.
It should be noted that the expansion into
kernels is not always guaranteed.  In particular, the separation of a
membrane kernel from a synaptic kernel may not always be possible if
the neuron is strongly nonlinear.  In such a case, an integral
approximation can still to be obtained and linearizations around
phase-locked solutions may be possible to examine stability.

We emphasize that we have computed the local
stability of the states.  This is in contrast to previous strategies
where the global stability is computed~\cite{herz,mirollo90,bottani}.
In these works, the class of initial conditions that will approach a
synchronous solution is found.  An absorption
criterion is used that assumes that once the oscillators are
within a given phase of each other then they remain locked and
no longer influence each other.  In our
formulation, the synaptic coupling can still
influence locked neurons
and be enough to destabilize the dynamics if certain other
criteria are not met.  In the case of excitatory coupling, we find
that although oscillators may approach each other globally they may
not be locally stable.
It had been previously assumed that the reasons
for disparate results for the stability of synchronized pulse-coupled
integrate-and-fire oscillators were a result of the singular
nature of the delta-distribution synaptic
coupling~\cite{hansel95,gerstner96}.  However,  it is
the assumptions regarding what constitutes a locked state 
that separates the conclusions.  Ideally, a combined analysis should be
done where a global stability analysis is used to examine the class of
initial 
conditions that approach a phase-locked solution
and then a local stability analysis is performed to see if the
locked state is sustainable.

We only considered heterogeneity in the applied currents which  is a
very simple form 
of disorder because it is not necessary to average over different
instances of the 
heterogeneity.  For instance if we choose $N$ random current values from
a fixed probability distribution and if $N$ is large enough, then a single
instance is equivalent to any other instance.  This is because the network
is insensitive to rearrangements of the current values between the
neurons.  We can always relabel the neurons in the network.
Only the distribution of the gaps between current values are
important and this is fixed for large enough $N$.  This property
allowed us to gain insight into the $N$ neuron network from the two
neuron network.
On the other hand if the
heterogeneity were to be included in the coupling strengths then
a single instance of the heterogeneity may not give an accurate
picture of what a typical random network will do.    
A rearrangement of the coupling strengths could greatly affect the
network dynamics.  Here, an understanding of two heterogeneously
coupled neurons may not give any insight into the behavior of a large
network. 
One must average over the heterogeneity to get an idea of what a typical
network will do.  Then one must probe the fluctuations around this
average. 
For two neurons it seems that
heterogeneity for synaptic strengths would have similar results to
what we derived for heterogeneity in the applied current. Phase-locked
solutions would be shifted away from the homogeneous solutions.  Our
stability results would then apply analogously.  However, the behavior
of large networks may be quite different from the behavior of two
neurons. 

\acknowledgements

I would like to thank N. Kopell for many clarifying discussions
and for constructive critiques of the manuscript.  
I thank
S. Epstein, B. Ermentrout, W. Gerstner, and J. White for many helpful 
suggestions.  I especially thank
J. Ritt for performing numerical simulations, generating figures,
and for his insightful comments on the manuscript.
 This work was supported by the National
Science Foundation grant DMS-9631755.

\appendix

\section{Conductance-based neuron model}\label{app:neu}

For numerical simulations we used a single-compartment model
neuron with 
inhibitory synapses obeying first-order kinetics~\cite{white97}.
Membrane potential in each point neuron obeyed the current balance
equation
\begin{equation}
C\frac{dV_i}{dt}=I_i-I_{Na}-I_K-I_L-I_s,
\label{membrane}
\end{equation}
where $C=1 \mu$F/cm$^2$, $I_i$ is the applied current,
$I_{Na}=g_{Na} m_\infty^3 h (V_i - V_{Na})$ and $I_K=g_Kn^4(V_i-V_K)$ are
the Hodgkin-Huxley type spike generating currents, $I_L= g_L(V_i-V_L)$
is the leak current and $I_s = \sum_j^N g_{s} S_j(t) (V_i-V_s)$ is the
synaptic current. 
The fixed parameters used were: $g_{Na}=30$~mS/cm$^2$, $g_K=20$~mS/cm$^2$, 
$g_L=0.1$~mS/cm$^2$,
$V_{Na}=45$~mV, $V_K=-75$~mV, $V_L=-60$~mV, $V_s=-75$~mV. 
The phenomena described here seem largely independent of specific
neuronal parameters.

The activation variable $m$ was assumed fast and substituted with
its asymptotic value
$m_\infty(v)=(1 + \exp[-0.08(v+26)])^{-1}$.  The inactivation
variable $h$ obeys
\begin{equation}
\frac{dh}{dt}=\frac{h_\infty(v)-h}{\tau_h(v)},
\end{equation}
with $h_\infty(v)=(1 + \exp[0.13(v+38)])^{-1}$, 
$\tau_h(v)=0.6/(1 + \exp[-0.12(v+67)])$.  The variable $n$ obeys
\begin{equation}
\frac{dn}{dt}=\frac{n_\infty(v)-n}{\tau_n(v)},
\end{equation}
with $n_\infty(v)=(1 + \exp[-0.045(v+10)])^{-1}$, 
$\tau_n(v)=0.5+2.0/(1+\exp[0.045(v-50)])$.

The gating variable $S_j$ for the synapse is assumed to obey first order
kinetics of the form
\begin{equation}
\frac{dS_j}{dt} = F(V_j)(1-S_j)-\beta S_j,
\label{synapse}
\end{equation}
where $F(V_j)=1/(1+\exp[-V_j])$, and $V_j$ is the voltage of the
pre-synaptic neuron and transmission delays are neglected.

The ODEs were integrated using a fourth-order Runge-Kutta
method.  The free parameters were scanned across
the following ranges:  for applied current $I_i$, 0-10 $\mu$A/cm$^2$; 
for $g_s$, 0-2~mS/cm$^2$; for the synaptic decay time constant $\tau_s$,
5-50~ms.

\section{Phase-locking of two coupled neurons with no memory}\label{app:two}
As an example, we consider a model where the response kernels
have no memory beyond the previous firing.  The phase-locked solutions
are then given by
\begin{eqnarray}
v_1(T+\phi_1 T)&=&1=I_1+\eta(T) +J \epsilon(T-\phi T),
\label{app:v1}\\ 
v_2(T+\phi_2 T)&=&1=I_2+\eta(T) +J \epsilon(\phi T) +J\epsilon(T+\phi T),
\label{app:v2},
\end{eqnarray}
This implies that $G(\phi)= J[\epsilon(\phi T)+\epsilon(T+\phi T)
-\epsilon(T-\phi T)]$.
The perturbed
dynamics Eq.~(\ref{mapping}) are given by $\delta(n)=M
\cdot \delta(n-1)$, where
\begin{equation}
\delta(n) = 
\left(\begin{array}{c} \delta_1(n) \\ \delta_2(n)
\end{array}
\right), \quad
M=\left(\begin{array}{cc} 
\frac{\dot{\eta}(lT)}{\dot{v}_1} &
\frac{J\dot{\epsilon}(lT-\phi T)}{\dot{v}_1} \\
\frac{J\dot{\epsilon}(lT+\phi T)}{\dot{v}_2} 
+\frac{J\dot{\epsilon}(\phi T) \dot{\eta}(lT)}{\dot{v}_1\dot{v}_2} &
\frac{\dot{\eta}(lT)}{\dot{v}_2} +
\frac{J^2\dot{\epsilon}(\phi T)\dot{\epsilon}(lT-\phi T)}{\dot{v}_1
\dot{v}_2} 
\end{array}
\right),
\label{app:M}
\end{equation}
\begin{eqnarray}
\dot{v}_1&=&\dot{\eta}(T)+J\dot{\epsilon}(T-\phi T), \label{v_1'}\\
\dot{v}_2&=&\dot{\eta}(T)+J\dot{\epsilon}(\phi T)+J\dot{\epsilon}(T+\phi
T),\label{v_2'} 
\end{eqnarray}
and the prime refers to the first derivative with respect to time. 
By definition, $v(t)$ must
approach the threshold from below which implies that $\dot{v}_i>0$.  

The eigenvalues can be solved for exactly.  We can verify that the
rows of $M(l)$ in  (\ref{app:M}) sum to unity and so one
eigenvalue is 1.  The 
other eigenvalue is then $\lambda={\rm Tr}\, M -1\equiv \det M$.
For stability, we require $|\lambda|<1$ which implies that $0<{\rm
Tr}\, M<2$ or $-1< \det M<1$,
where
\begin{equation}
{\rm Tr}\, M = \frac{\dot{\eta}(T)}{\dot{v}_1} +
\frac{\dot{\eta}(T)}{\dot{v}_2} + 
J^2\frac{\dot{\epsilon}(\phi T)\dot{\epsilon}(T-\phi T)}{\dot{v}_1\dot{v}_2},
\label{trace}
\end{equation}
\begin{equation}
\det M =
\frac{\dot{\eta}(T)}{\dot{v}_1\dot{v}_2}-J^2\frac{\dot{\epsilon}(T-\phi
T)\dot{\epsilon}(T+\phi T)}{\dot{v}_1\dot{v}_2}.
\label{det}
\end{equation}

We now apply our results of Sec.~\ref{sec:stab} and compare to the exact
eigenvalues. If all the elements of $M$ are positive then we immediately
see that stability is achieved. Since the rows must
sum to 1 and the elements are positive then each element itself is
less than 1.  This immediately implies that  $0<\det M<1$ as required
for stability.  However, this condition although sufficient
for stability is certainly not necessary.  An examination of  ${\rm
Tr}\, M$ and $\det M$  shows that the elements of $M$ can become
negative and still maintain $0<{\rm Tr}\, M<2$ required for stability.

We next examine the necessary condition $G'(\phi)>0$ of Theorem
\ref{nec}.  We first note 
that $G'(\phi)=JT[\dot{\epsilon}(\phi T) +\epsilon_1'(T+\phi
T)+\dot{\epsilon}(T-\phi T)]$. Consider the synchronous solution
for synchrony ($\phi=0$) where
$G'(0)=JT[\dot{\epsilon}(0)+2\dot{\epsilon}(T)] >0$ is clearly
necessary for stability.  If in addition, $\dot{\epsilon}(0)=0$ and
$\dot{\eta}(T)>0$ then an inspection of condition (\ref{trace}) shows
that $G'(0)>0$ is necessary and sufficient for stability.
For the anti-synchronous solution ($\phi=0.5$), $G'(0.5)=
JT[2\dot{\epsilon}(0.5T)+\dot{\epsilon}(1.5T)]>0$ is necessary for
stability from
condition (\ref{trace}) or (\ref{det}).
However, unlike the synchronous case, it cannot be made into a
sufficient condition without making additional nontrivial constraints
on $\dot{\epsilon}(0.5T)$ and $\dot{\epsilon}(1.5T)$.

\end{document}